\documentclass[prd,nofootinbib]{revtex4-2}
\usepackage{amsmath, amssymb,subfigure} 
\usepackage{graphicx}
\usepackage{multirow}
\usepackage{hyperref}

\begin{document}

\title{Neutron stars in 4D Einstein-Gauss-Bonnet gravity}
\author{Alejandro Saavedra} 
\email{alsaavedra2019@udec.cl} 
\affiliation{Departamento de Física, Universidad de Concepción,  
Casilla 160-C, Concepción, Chile}

\author{Octavio Fierro} 
\email{ofierro27@yahoo.com} 
\affiliation{Facultad de Ingenier\'{i}a, Arquitectura y Diseño, Universidad San Sebasti\'{a}n,
Lientur 1457, Concepción 4080871, Chile}

\author{Michael Gammon} 
\email{gammon@uwaterloo.ca} 
\affiliation{Department of Physics and Astronomy,
University of Waterloo, Waterloo, Ontario, Canada, N2L 3G1}

\author{Robert B. Mann} 
\email{rbmann@uwaterloo.ca} 
\affiliation{Department of Physics and Astronomy,
University of Waterloo, Waterloo, Ontario, Canada, N2L 3G1}

\author{Guillermo Rubilar} 
\email{grubilar@udec.cl} 
\affiliation{Departamento de Física, Universidad de Concepción,  
Casilla 160-C, Concepción, Chile}

\begin{abstract}
    \vskip1cm\noindent
    Since the derivation of a well-defined $D\to4$ limit for 4D Einstein-Gauss-Bonnet (4DEGB) gravity coupled to a scalar field, there has been considerable interest in testing it as an alternative to Einstein's general theory of relativity. Past work has shown that this theory hosts interesting compact star solutions which are smaller in radius than a Schwarzschild black hole of the same mass in general relativity (GR), though the stability of such objects has been subject to question. In this paper we solve the equations for radial perturbations of neutron stars in the 4DEGB theory with Skyrme Lyon (SLy)/Brussels-Montreal Skyrme functionals (BSk) class equations of state (EOSs), along with the Müller-Serot (MS2) EOS, and show that the coincidence of stability and maximum mass points in GR is still present in this modified theory, with the interesting additional feature of solutions reapproaching stability near the black hole solution on the mass-radius diagram. Besides this, as expected from previous work, we find that larger values of the 4DEGB coupling $\alpha$ tend to increase the mass of neutron stars of the same radius (due to a larger $\alpha$ weakening gravity) and move the maximum mass points of the solution branches closer to the black hole horizon.
    
    \end{abstract}

\maketitle

\section{Introduction}

Despite the empirical success and predictive power of Einstein's general theory of relativity (GR), modified theories of gravity remain interesting as a possible way to address issues in modern cosmology \cite{bueno2016,sotiriou_2010,nojiribook,clifton_2012}, quantize gravity \cite{ahmed2017,stelle1977}, and eliminate spacetime singularities \cite{brandenberger1992nonsingular,BRANDENBERGER_1993,brandenberger1995implementing}. Perhaps even more important is the need for phenomenological competitors against which GR can be tested in the most stringent manner possible. Early attempts at modifying Einsteinian gravity can be traced back to work by  Weyl \cite{weyl1921} and Eddington \cite{eddington1921}, and continue through to present day research.

Amongst the plethora of modified gravitational theories, higher curvature theories are among the most popular. In such theories the assumed linear relationship between curvature and energy-momentum found in GR is replaced with a relation that  depends on an arbitrary sum of powers of the curvature tensor. Lovelock theories \cite{lovelock1971} have long been the preferred class of  higher curvature theories since the resultant field equations are of second order.

Lovelock theories of gravity have historically been treated as mathematical curiosities since Lovelock's theorem ensures that the higher order terms vanish identically in four spacetime dimensions or less ($D\leq4$). The first of these higher order terms -- the Gauss Bonnet (GB) term -- is  quadratic in the curvature,
\begin{equation}\label{eq:gbaction}
\begin{aligned}
S_{D}^{G B} &= \alpha \int d^{D} x \sqrt{-g} \left[R^{\mu\nu\rho\tau}R_{\mu\nu\rho\tau} - 4 R^{\mu\nu}R_{\mu\nu} + R^2 \right] \\
&\equiv \alpha \int d^{D} x \sqrt{-g} \mathcal{G},
\end{aligned}
\end{equation}
and is the integral of a total derivative in $D=4$, thus contributing nothing to the system's dynamics. However it was recently shown 
\cite{glavanlin}  that the Gauss-Bonnet contribution to solutions to the $D$-dimensional field equations can be nontrivial for $D\to 4$
under the following rescaling of the GB coupling constant:
\begin{equation}\label{eq:alpharescale}
	\alpha \rightarrow \frac{\alpha}{D-4}.
\end{equation}
Despite this apparent  violation of the Lovelock theorem (which states that GR with cosmological constant is the most general theory of gravity in 3+1 dimensions that is wholly described by a metric and second order field equations), a number of sensible four-dimensional metrics can be obtained. This was done for spherical black holes \cite{glavanlin,kumar2022,fernandes2020,kumar2020,kumar2022_2}, cosmological solutions \cite{glavanlin,li2020,kobayashi2020}, starlike solutions \cite{doneva2021,Charmousis2022}, radiating solutions \cite{ghosh2020}, collapsing solutions \cite{malafarina2020}, etc, with all of these solutions carrying imprints of higher curvature corrections inherited from  their $D > 4$ counterparts. 

In reality the existence of  such limiting solutions  does not actually imply the existence of a well-defined 4D theory, and a number of objections in this vein quickly appeared \cite{gurses2020,ai2020,shu2020}. This shortcoming was addressed by two independent groups \cite{Hennigar2020,Fernandes:2020nbq}, both of which derived consistent versions of what has come to be known as 4D Einstein-Gauss-Bonnet (4DEGB) gravity, making use of the same rescaling \eqref{eq:alpharescale} first introduced by Glavan and Lin \cite{glavanlin}. In both cases a scalar field is introduced into the action thus preserving the Lovelock theorem and making 4DEGB gravity a Horndeski theory of gravity. In the former this is done via a conformal rescaling trick (analogous to an earlier procedure wherein the $D\to2$ limit of GR was obtained \cite{mannross2d}), and in the latter a Kaluza-Klein dimensional reduction technique \cite{Lu:2020iav}. These two approaches yield (up to trivial field redefinitions)  identical theories, with the exception that the latter method yields additional terms in the metric field equations that depend on the curvature of the maximally symmetric $(D-4)$-dimensional space. Taking these terms to vanish yields the 4DEGB action term:
\begin{equation}\label{eq:4DEGBactionterm}
\begin{aligned}
S_{4}^{GB}
=&\alpha \int d^{4} x \sqrt{-g}\left[  \phi \mathcal{G}+4 G_{\mu \nu} \nabla^\mu \phi \nabla^\nu \phi-4(\nabla \phi)^2 \square \phi+2(\nabla \phi)^4\right],
\end{aligned}
\end{equation}
 where $\phi$ is a new scalar field.
 
The above contribution is added to the usual Einstein-Hilbert term in the full theory and acts as a modification to standard GR. Surprisingly, the static, spherically symmetric black hole solutions to the resultant field equations match those following from the naive $D \rightarrow 4$ limit of 
$D>4$ solutions found in \cite{glavanlin} without ever actually referencing a higher dimensional spacetime. This full theory has been shown to be an interesting phenomenological competitor to GR \cite{Clifton:2020xhc,Charmousis2022,Zanoletti:2023ori}. 
Despite much exploration \cite{Fernandes2022}, the role played by these
higher curvature terms in real gravitational dynamics is still not fully understood. One important arena for testing such theories against standard general relativity is via observations of compact astrophysical objects like neutron stars. The correct theory should be able to accurately describe recent and future gravitational wave observations of astrophysical objects existing in the mass gap between the heaviest compact stars and the lightest black holes. 

Modern observational astrophysics is rich in its detection of compact objects and as such our understanding of highly dense gravitational objects is rapidly advancing. However, there is as of yet no strong consensus on their underlying physics. A number of such objects have been recently observed that are inconsistent with standard GR and a simple neutron star equation of state.  Recently it was shown that the secondary component of the merger GW190814 is well described as a slowly rotating neutron star (NS) in the 4DEGB theory without resorting to exotic equations of state (EOSs), while also demonstrating that the equilibrium sequence of neutron stars asymptotically matches the minimum mass black hole solution, thus closing the mass gap between NS/black holes of the same radius \cite{Charmousis2022}. 

In a similar vein, investigations on quark stars in the 4DEGB theory have shown promise for describing unusual astronomical objects like HESS J1731-347 \cite{Horvath_2023}, PSR J0030+0451 \cite{Miller_2019}, PSR J0740+6620 \cite{salmi2024}, and GW190814 \cite{Abbott_2020}. In these cases the asymptotic solutions mentioned above for neutron stars were seen again for quark stars, and a comparison to a modified version of the Buchdahl bound \cite{CHAKRABORTY2020,gammon2024} for 4DEGB gravity was carried out \cite{Gammon2023,gammon2024}. It was found that in all cases, in the limit of large central pressure, the compact star solutions asymptotically approached the minimum mass black hole of the theory (while also smoothly joining with the minimum mass point of the modified Buchdahl bound) in the limit of large central pressure. Such solutions beg the question of stability - in general relativity an uncharged, spherical, gravitating body is stable in the part of the solution branch where $\partial M/ \partial \rho_c \geq 0$ \cite{Glendenning} (i.e., the transition to instability happens at the maximum mass point). Once a charge is introduced into the system, this coincidence no longer holds \cite{arbanil2015,goncalves2020,zhang2021} and the point at which the fundamental mode eigenfrequency $\omega_0^2$ becomes negative is offset from the maximum mass point. In the literature thus far it has been unclear whether including nonzero coupling to the 4DEGB theory will have a similar offsetting effect, and whether the parts of the solution curves corresponding to extreme compact objects (or ECOs) could exist in a universe described by the 4DEGB theory. If there are indeed stable solutions that approach the black hole horizon, then for that set of parameters we can expect a universe in which the radii of compact stars can be arbitrarily close to the horizon size of black holes.

This naturally also invokes curiosity about the gravitational collapse scenario in 4DEBG gravity, leaving an interesting avenue for future work.

In this article we tackle the former problem, investigating the stability of neutron star solutions in 4DEGB gravity for the Skyrme Lyon (SLy) EOS, the Brussels-Montreal Skyrme functionals (BSk) family of EOSs, and the Müller-Serot  EOS (MS2). We found that the qualitative differences between the resultant mass-radius curves for these different EOSs were minimal, and that increasing the 4DEGB coupling constant $\alpha$ had the effect of generally inflating the mass-radius (MR) profiles. In the case of the MS2 EOS we see maximum mass points which are offset further from the intersection point of the solution curve with the black hole horizon than in the nonrelativistic cases. Furthermore, we find numerically that the transition from stability to instability in the 4DEGB theory is still coincident with the maximum mass point of the MR curves. Interestingly, the MR solution curves, after reaching the maximum mass, transit through a region of unstable solutions that tend to reapproach stability at the limit of large central density (i.e., the solutions closest to the black hole for a given $\alpha$), possibly hinting at black hole-sized stable ECOs as a general feature of this theory. Even ignoring these hypothetical objects, for large enough $\alpha$ the solution curves do not reach a maximum mass until they merge with the black hole horizon. This implies that the theory does predict the existence of objects  that are smaller than the Schwarzschild radius in GR yet are stable against radial perturbations. The caveat here is that our nonrelativistic equations of state sometimes exceed the causality limit (i.e., $v_{\rm sound} = c$) before this maximum mass point is reached. Such solutions should not be considered physical, although analogous behavior is also observed for the MS2 EOS which does take relativistic considerations into account. In the latter case no such issue with causality is present, and the full solution space for all EOSs is included for comparison.

The outline of our paper is as follows: in Sec. \ref{sec:4degb} we introduce the formalism and field equations of 4DEGB gravity, including an overview of the spherically symmetric black hole solution which matches that found in \cite{glavanlin}. In Sec. \ref{sec:ns} a perfect fluid energy-momentum tensor is employed and the modified Tolman-Oppenheimer-Volkoff (TOV) equations are derived for 4DEGB gravity. After this, Sec. \ref{sec:eos} introduces the SLy/BSk/MS2 neutron star EOSs which are implemented in our numerical calculations. We use the section to examine the sound speed inside stars described by each EOS as a function of density, and to show the resultant MR curves from solving the modified TOV equations alongside these EOSs. This is followed by a thorough study of the stability under adiabatic radial oscillations of these neutron stars solutions in Sec. \ref{sec:radosc}. Finally, in Sec. \ref{sec:conc} we summarize our results and discuss interesting avenues for future research.

\section{4D Einstein-Gauss-Bonnet Gravity}\label{sec:4degb}

The 4DEGB theory is defined by adding \eqref{eq:4DEGBactionterm} to the Einstein-Hilbert action \cite{Hennigar2020}:
\begin{equation}
\label{eq:EGB-gravity}
  S_{\rm EGB} = \frac{1}{2\kappa} \int d^4 x \sqrt{-g} \left[ R - 2 \Lambda + \alpha \left(\phi \mathcal{G} + 4 G_{\mu\nu} \nabla^{\mu} \phi \nabla^{\nu} \phi - 4 (\nabla \phi)^2 \square \phi + 2 ((\nabla \phi)^2)^2 \right) \right] + S_{\rm m},     
\end{equation} 
where $\kappa = 8\pi Gc^{-4}$, $\alpha$ is the 4DEGB coupling constant (with units of length squared), $\phi$ is the (dimensionless)  scalar field, $S_{\rm m}$ is the matter action and 
\begin{equation}
\mathcal{G} = R_{\mu\nu\rho\sigma} R^{\mu\nu\rho\sigma} - 4 R_{\mu\nu} R^{\mu\nu} + R^2    
\end{equation}
is the Gauss-Bonnet term. We note that an important property of the action \eqref{eq:EGB-gravity} is its shift symmetry in the scalar field, i.e. it remains invariant under the transformation
\begin{equation}
    \phi \rightarrow \phi + \mathcal{C}, \label{eq:shift-symmetry}
\end{equation}
for a constant $\mathcal{C}$.

The field equation for the scalar field is given by \cite{Hennigar2020}
\begin{align}
\label{eq:scalar-FEq}
     &\mathcal{G} - 8 G_{\mu\nu} \nabla^{\mu} \nabla^{\nu} \phi - 8 R_{\mu\nu} \nabla^{\mu}\phi\nabla^{\nu}\phi + 8 (\square \phi)^2  -8 \nabla_{\mu}\nabla_{\nu} \phi \nabla^{\mu}\nabla^{\nu} \phi  -16 \nabla_{\mu} \nabla_{\nu} \phi \nabla^{\nu} \phi \nabla^{\mu} \phi\nonumber \\
    &\quad  - 8 (\nabla \phi)^2 \square \phi  = 0,
\end{align}
while variation of the action with respect to the metric leads to the following field equations
\begin{align}
\label{eq:metric-FEq}
   & G_{\mu\nu} + \Lambda g_{\mu\nu} + \alpha \left[\phi H_{\mu\nu} -2R [\nabla_{\mu}\nabla_{\nu}\phi + (\nabla_{\mu}\phi)(\nabla_{\nu}\phi)] + 8 R^{\rho}_{\ (\mu} \nabla_{\nu)}\nabla_{\rho}\phi + 8 R^{\rho}_{\ (\mu}\nabla_{\nu)}\phi\nabla_{\rho} \phi \right. \nonumber \\
    &\quad -2G_{\mu\nu} [(\nabla \phi)^2 + 2\square \phi] - 4[\nabla_{\mu}\nabla_{\nu} \phi + (\nabla_{\mu}\phi)(\nabla_{\nu}\phi)]\square\phi - [g_{\mu\nu}(\nabla \phi)^2 - 4 (\nabla_{\mu}\phi)(\nabla_{\nu}\phi)] (\nabla \phi)^2 \nonumber\\
    &\quad + 8 \nabla_{\rho}\nabla_{(\mu}\phi(\nabla_{\nu)} \phi) \nabla^{\rho} \phi - 4 g_{\mu\nu} R^{\rho\sigma} [\nabla_{\sigma}\nabla_{\rho}\phi + (\nabla_{\sigma}\phi)(\nabla_{\rho} \phi)] + 2g_{\mu\nu} (\square \phi)^2 \nonumber \\
    &\quad -2g_{\mu\nu} (\nabla_{\rho} \nabla_{\sigma}\phi)( \nabla^{\rho} \nabla^{\sigma}\phi) - 4g_{\mu\nu} (\nabla_{\rho}\nabla_{\sigma}\phi) (\nabla^{\rho}\phi)(\nabla^{\sigma} \phi) + 4 (\nabla_{\mu}\nabla_{\rho}\phi)(\nabla_{\nu}\nabla^{\rho}\phi) \nonumber\\
    &\quad + 4 R_{\mu\rho\nu\sigma} [\nabla^{\rho}\nabla^{\sigma}\phi + (\nabla^{\rho}\phi)(\nabla^{\sigma} \phi)] = \frac{8\pi G}{c^4} T_{\mu\nu}, 
\end{align}    
where 
\begin{equation}
     T_{\mu\nu} := - \frac{2}{\sqrt{-g}} \frac{\delta S_{\rm m}}{\delta g^{\mu\nu}},
\end{equation}
and 
\begin{equation}
H_{\mu\nu} =2 R_{\mu\rho\sigma \lambda} R_{\nu}^{\ \rho\sigma \lambda} -4R_{\mu\rho} R_{\nu}^{\ \rho} - 4R_{\mu\rho\nu\sigma} R^{\rho\sigma} + 2 R R_{\mu\nu}    - \frac{1}{2} \mathcal{G} g_{\mu\nu},
\end{equation}
which identically vanishes in four dimensions and less. 

In what follows we will work with a cosmological constant equal to zero.

\subsection{Black hole solution}

This theory possesses an exact vacuum solution, with a line element given by
\begin{equation}
\label{eq:BH-metric}
    ds^2 = - f(r) (c dt)^2 +  \frac{dr^2}{f(r)} + r^2 (d\theta^2 + \sin^2\theta d\varphi^2), 
\end{equation}
where the metric function $f(r)$ and the (derivative of the) scalar field $\phi$ are given by  \cite{Hennigar2020}
\begin{gather}
    f(r) = 1 + \frac{r^2}{2\alpha} \left( 1 - \sqrt{1 + \frac{8\alpha GM}{c^2 r^3}}\right), \label{eq:f-exterior}\\
    \frac{d\phi}{dr} = \frac{\sqrt{f} - 1}{r \sqrt{f}}, \label{eq:SF-exterior}
\end{gather}
and $M$ is an integration constant. This solution is asymptotically flat, and we can interpret $M$ as the mass of a nonrotating black hole. There are two horizons for $\alpha < 0$, as well as for $M > c^2\sqrt{\alpha}/G = M_{\rm min}$ if $\alpha > 0$. The outer event horizon \cite{Charmousis2022} is located at 
\begin{equation} \label{eq:BH-horizon}
    R_{\rm h} = \frac{GM}{c^2} + \sqrt{\frac{G^2M^2}{c^4} - \alpha}
\end{equation}
which for $\alpha>0$ is smaller than its Schwarzschild counterpart.
There are other branches of spherically symmetric solutions in this theory, but this one is the only asymptotically flat spherically symmetric solution that is free of naked singularities  \cite{Fernandez2021}. For this reason, the spacetime outside a spherically symmetric neutron star will be given by the line element \eqref{eq:BH-metric} with metric function \eqref{eq:f-exterior}. Writing $ f(r)=1+2\varphi(r)/c^2$, we can   compute the gravitational force per unit of mass in 4DEGB due to a spherical body,
\begin{equation}\label{force}
    \vec{f} = -\frac{d\varphi}{dr}\hat{r} = -\frac{c^2r}{2\alpha}
\left(1- \frac{c^2r^3+2 \alpha G M}{c^2r^3+8\alpha G M}\sqrt{1+\frac{8 \alpha G M}{c^2 r^3}}\right)  \hat{r},
\end{equation}
which is smaller in magnitude than its Newtonian $\alpha=0$ counterpart ($\vec{f}_N = - GM \hat{r}/r^2$) for $\alpha>0$. The expression in  \eqref{force} vanishes at  $r=(\alpha GM/c^2)^{1/3}$, but this is always at a smaller value of $r$ than the outer horizon  [as defined in \eqref{eq:BH-horizon}] of the corresponding black hole. Hence, the gravitational force outside of any spherical body, while weaker than in GR, is always attractive provided $\alpha>0$.  If $\alpha<0$ then the corresponding gravitational force is more attractive than in GR. However, the requirement that atomic nuclei should not be shielded by a horizon yields the empirical constraint  \cite{Charmousis2022}
\begin{equation}
    \alpha \gtrsim - 10^{-30} \ {\rm m^2},
\end{equation}
 making the associated gravitational effects totally undetectable. For practical purposes we can exclude negative $\alpha$ from our analysis. 
 
An upper bound for the coupling constant 
\begin{equation}
    0 < \alpha \lesssim 10^{10} \ {\rm m^2}
\end{equation}
has been found using LAGEOS satellites \cite{Fernandes2022}. Inclusion of preliminary calculations on recent GW data suggests that these constraints could be even tighter  \cite{Fernandes2022},
\begin{equation}
    0 < \alpha \lesssim 10^{7} \ {\rm m^2}
\end{equation}
though a proper calculation remains to be carried out.

\section{Neutron star solutions}\label{sec:ns}

We consider neutron star configurations as a static, spherically symmetric perfect fluid in hydrostatic equilibrium. The spacetime can be described by the line element \begin{equation}
\label{eq:static-metric}
    ds^2 = - e^{\chi(r)} f(r) (c dt)^2 +  \frac{dr^2}{f(r)} + r^2 (d\theta^2 + \sin^2\theta d\varphi^2), 
\end{equation}
where $f(r)$ and $\chi(r)$ are  metric functions. The matter inside the star will be described in terms of the energy-momentum tensor of a perfect fluid, given by
\begin{equation}
    T_{\mu\nu} = \left(\epsilon + P \right) \frac{u_{\mu}}{c} \frac{u_{\nu}}{c} + P g_{\mu\nu},
\end{equation}
where $\epsilon = \rho c^2$ is the energy density (and $\rho$ the mass density), $P$ is the isotropic pressure, $u^{\mu} = dx^{\mu}/d\tau$ is the four-velocity of a fluid element, and $\tau$ is the proper time.

Since we are considering the static case, the only nonvanishing component of the four-velocity is $u_0$ so that $u_{\mu} = (u_0,0,0,0)$. Using the identity $g_{\mu\nu} u^{\mu} u^{\nu} \equiv -c^2$ we can obtain 
\begin{equation}
    u_0 = - c \sqrt{f} e^{\chi/2}.
\end{equation}

The energy-momentum tensor is then given by
\begin{equation}
\label{eq:Tmunu-static}
T_{\mu\nu} = {\rm diag}(\epsilon e^{\chi} f, P/f, Pr^2, Pr^2\sin^2\theta). 
\end{equation}

Inserting the metric \eqref{eq:static-metric} and the components of the energy-momentum tensor \eqref{eq:Tmunu-static} into the field equations  \eqref{eq:metric-FEq}, we obtain  Eqs. \eqref{eq:tt-static} and \eqref{eq:rr-static} in Appendix \ref{appFE}. 

Since the action is invariant under the transformation \eqref{eq:shift-symmetry}, it can be shown that Eq. \eqref{eq:scalar-FEq} for the scalar field in the spacetime given by \eqref{eq:static-metric} can be recast as \cite{Saravani2019}
\begin{equation}
\label{eq:SF-Current}
    \frac{dj^{r}}{dr} = 0, 
\end{equation}
with 
\begin{equation}
     j^{r} = \frac{4\alpha}{r^2}[(r\phi' - 1)^2 f - 1][2\phi' f - \chi' f  - f'].
\end{equation}

For the spacetime to be asymptotically flat and match the exterior solution implies
\begin{equation}
    (r\phi' - 1)^2 f - 1 = 0
\end{equation}
the solution of which is given by  \eqref{eq:SF-exterior}, which also holds for the interior of the neutron star. 

Replacing the expression \eqref{eq:SF-exterior} for the scalar field in Eqs. \eqref{eq:tt-static} and \eqref{eq:rr-static}, and solving for the derivatives of $f$ and $\chi$ yields
\begin{gather}
\frac{df}{dr} = - \frac{(8\pi G/c^4) \epsilon  r^{4} + \alpha f^{2}  + (r^{2} -2\alpha) f  -  r^{2}+\alpha }{ r \left(r^2 - 2\alpha f + 2 \alpha \right)}, \label{eq:df-static} \\
\frac{d\chi}{dr} =  \frac{8\pi G}{c^4} \frac{r^{3} \left(\epsilon + P\right)}{f  \left(r^2 - 2\alpha f  + 2\alpha  \right)} \label{eq:dchi-static}.
\end{gather}

The only nontrivial component of the conservation equation 
$\nabla_{\mu} T^{\mu r} = 0$ for the energy-momentum tensor \eqref{eq:Tmunu-static} is the $r$ component, which gives 
\begin{equation}
\label{eq:DTmunu-r}
\frac{dP}{dr} = - \frac{1}{2} (\epsilon + P) \left( \chi' + \frac{f'}{f}\right). 
\end{equation}

Inserting Eqs. \eqref{eq:df-static} and \eqref{eq:dchi-static} in \eqref{eq:DTmunu-r}, we find 
\begin{equation}
\frac{dP}{dr} = - \frac{\left(\epsilon + P\right) \left[-\alpha f^{2} - \left(r^{2}-2 \alpha \right) f + (8\pi G/c^4) r^{4} P+ r^{2}-\alpha \right]}{2  r f \left(r^{2}-2 \alpha f + 2 \alpha \right)}, \label{eq:dP-static}
\end{equation}
which defines the modified TOV equation for this theory in terms of the metric function $f$, the pressure, and the energy density.

We determine the boundary conditions at the origin by assuming that each function is regular near $r=0$, that is,
\begin{equation}
    h(r) = h_0 + h_1 r + h_2 r^2 + \cdots, \quad r \to 0, \label{eq:Taylor-expansion}
\end{equation}
with $h = \{f,\chi, \phi,\epsilon,P\}$. Replacing this expansion in Eqs. \eqref{eq:SF-exterior}, \eqref{eq:df-static}, \eqref{eq:dchi-static}, and \eqref{eq:dP-static}, solving order by order, and requiring that the scalar field is regular at the origin implies 
\begin{gather}\label{eq:frexp}
f(r) = 1 + \frac{1}{2\alpha} \left(1 - \sqrt{1 + \frac{32\pi G}{3c^4} \alpha \epsilon_{\rm c} } \right) r^2 + \mathcal{O}(r^3)\\
    \chi(r) = \chi_{\rm c} + \frac{4 \pi G}{c^4} \frac{\epsilon_{\rm c} + P_{\rm c}}{\sqrt{1 + (32 \pi G/3c^4) \alpha \epsilon_{\rm c} }} r^2 + \mathcal{O}(r^3),\\
    P(r) = P_{\rm c} +  \mathcal{O}(r^2)
\end{gather}
where $P_{\rm c}$ is the central pressure, $\epsilon_{\rm c}$ is the central energy density, and $\chi_{\rm c}$ is an arbitrary constant that is fixed by the matching with the exterior solution. To numerically integrate the system of differential equations it is necessary that the value of $\chi$ at the origin is known, and for this we will take advantage of the fact that the differential equation \eqref{eq:dchi-static} is linear in $\chi$. We write 
\begin{equation}
    \chi(r) = \chi_{\rm num}(r) + \chi_{\rm c},
\end{equation}
where $\chi_{\rm num}$ is the function that will be integrated, with $\chi_{\rm num}(0) = 0$ and the constant $\chi_{\rm c}$ obtained by imposing $\chi(R) = 0$.

\section{Equations of state}\label{sec:eos}

\begin{figure}[!t]
    \centering
    \includegraphics[width=0.7\linewidth]{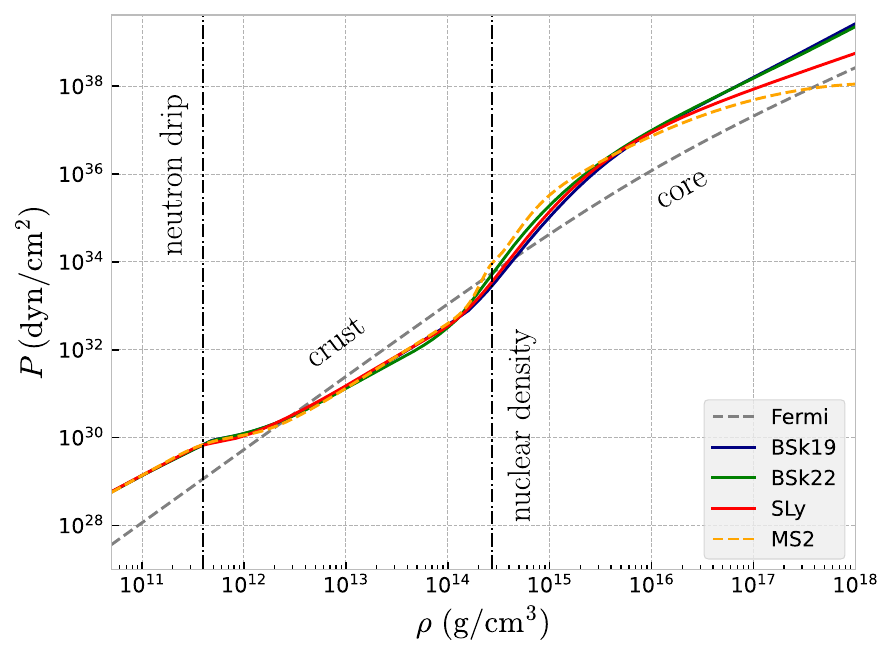} 
    \caption{Pressure-density relation for the SLy EOS (red solid line), BSk19-22 EOSs (blue and green solid lines), MS2 EOS (orange dashed line), and the Fermi EOS (gray dashed line) used by Oppenheimer and Volkoff in Ref. \cite{Oppenheimer39}. The vertical dash-dotted lines mark the three main regimes inside a neutron star. Descending in density, above the nuclear density ($\rho_{\rm n} = 2.8 \times 10^{14} \ {\rm g/cm^3}$)   is the core of the NS, the inner crust is below the nuclear density, but above the neutron drip ($\rho_{\rm drip} = 4 \times 10^{11} \ {\rm g/cm^3}$), and the outer crust below the neutron drip.}
    \label{fig:EoS}
\end{figure}

In order to solve the system of equations given by \eqref{eq:df-static}-\eqref{eq:dP-static}, we also need an EOS relating the star's internal pressure with its density. We will use realistic EOSs to describe all regions inside a neutron star. In particular, we consider the Skyrme Lyon (SLy) \cite{Haensel2004} EOS, previously used in \cite{Charmousis2022}, along with the Brussels-Montreal Skyrme functionals (BSk) family of EOSs \cite{Potekhin2013,Pearson2019}, which incorporate refinements to the SLy EOS that improve the fit to both nuclear matter properties and neutron star observations. We note that these equations of state  are all nonrelativistic; although it is common practice in the literature to consider the corresponding solutions up to densities that respect the causality condition \cite{Charmousis2022,lattimer2021neutron,lin2022realistic}, technically such solutions can only be trusted when the speed of sound is a small fraction of the speed of light. For this reason we also consider the MS2 (Müller and Serot) \cite{Muther1996} equation of state which does include relativistic considerations in the derivation. However, since the pressure-density relation of MS2 is extremely similar to its nonrelativistic cousins in all but the very highest density regime $\sim 10^{16} \ \textrm{g/cm}^3$ (see Fig. \ref{fig:EoS}) where the neutron star solutions are already unstable, the relevant physical results are quite similar and useful for comparison. 

To start, the SLy EOS has the following analytic parametrization:
\begin{equation} \label{eq:SLy-EoS}
    \begin{aligned}
        \zeta &= \frac{a_1 + a_2\xi + a_3 \xi^3}{1 + a_4\xi} f_0(a_5(\xi - a_6)) + (a_7 + a_8\xi) f_0(a_9(a_{10} - \xi)) \\
        &\quad + (a_{11} + a_{12} \xi) f_0(a_{13}(a_{14} - \xi)) + (a_{15} + a_{16} \xi) f_0(a_{17} (a_{18} - \xi)),
    \end{aligned}
\end{equation}
where $\zeta = \log_{10}(P/{\rm dyn \ cm^{-2}})$, $\xi = \log_{10}(\rho/{\rm g \ cm^{-3}})$, $f_0(x) = 1/(1 + e^{x})$, and $\{a_1,\dots,a_{18}\}$ are dimensionless coefficients whose values can be found in Table \ref{tab:SLy-parameter} of  Appendix \ref{app:EoS}.

The analytic parametrization of the second family (BSk) of EOSs is given by
\begin{equation} \label{eq:BSk-EoS}
    \begin{aligned}
        \zeta &= \frac{a_1 + a_2\xi + a_3 \xi^3}{1 + a_4\xi} f_0(a_5(\xi - a_6)) + (a_7 + a_8\xi) f_0(a_9(a_{6} - \xi)) \\
        &\quad + (a_{10} + a_{11} \xi) f_0(a_{12}(a_{13} - \xi)) + (a_{14} + a_{15} \xi) f_0(a_{16} (a_{17} - \xi)) \\
        &\quad + \frac{a_{18}}{1 + [a_{19}(\xi - a_{20})]^2} + \frac{a_{21}}{1 + [a_{22}(\xi - a_{23})]^2}.
    \end{aligned}
\end{equation}
The value of the coefficients $\{a_1,\dots,a_{23}\}$ can be found in  Table \ref{tab:BSk-parameter} of  Appendix \ref{app:EoS}.

Finally, the MS2 equation of state can be parametrized by \cite{Gungor2011}
\begin{equation} \label{eq:MS2-EoS}
    \zeta = \zeta_{\rm low} f_0(a_1(\xi - c_{11})) + f_0(a_2(c_{12} - \xi)) \zeta_{\rm high},
\end{equation}
where 
\begin{equation}
    \zeta_{\rm low} = [c_1 + c_2(\xi - c_3)^{c_4}]f_0(c_5(\xi - c_6)) + (c_7 + c_8\xi) f_0(c_9(c_{10} - \xi))
\end{equation}
and
\begin{equation}
    \zeta_{\rm high} = (a_3 + a_4\xi)f_0(a_5(a_6 - \xi)) + (a_7 + a_8\xi + a_9 \xi^2) f_0(a_{10}(a_{11} - \xi))
\end{equation}
describe the low and high density regimes, respectively. The (dimensionless) coefficients $\{c_1,\dots,c_{12}\}$ and $\{a_1,\dots,a_{11}\}$ for the MS2 EOS are tabulated in Table \ref{tab:MS2-parameter} of  Appendix \ref{app:EoS}.

An important quantity that characterizes the relationship between pressure and density under adiabatic conditions is the adiabatic index, denoted by $\Gamma$ and defined as follows \cite{MTW}:
\begin{equation}
    \Gamma  := \frac{\epsilon + P}{P} \frac{dP}{d\epsilon} .
\end{equation}

As usual, the speed of sound is defined by 
\begin{equation}
    v_{\rm s} := \sqrt{\frac{dP}{d\rho}}
\end{equation}
and at the very least must be less or equal to the speed of light so that causality is not violated. Ideally for nonrelativistic equations of state we would only consider sound speeds much less than $c$, though due to the similarity to the relativistic MS2 EOS in the relevant density regime we keep the higher sound speed solutions in all cases for comparison.

\begin{figure}
    \centering
    \includegraphics[width=0.7\linewidth]{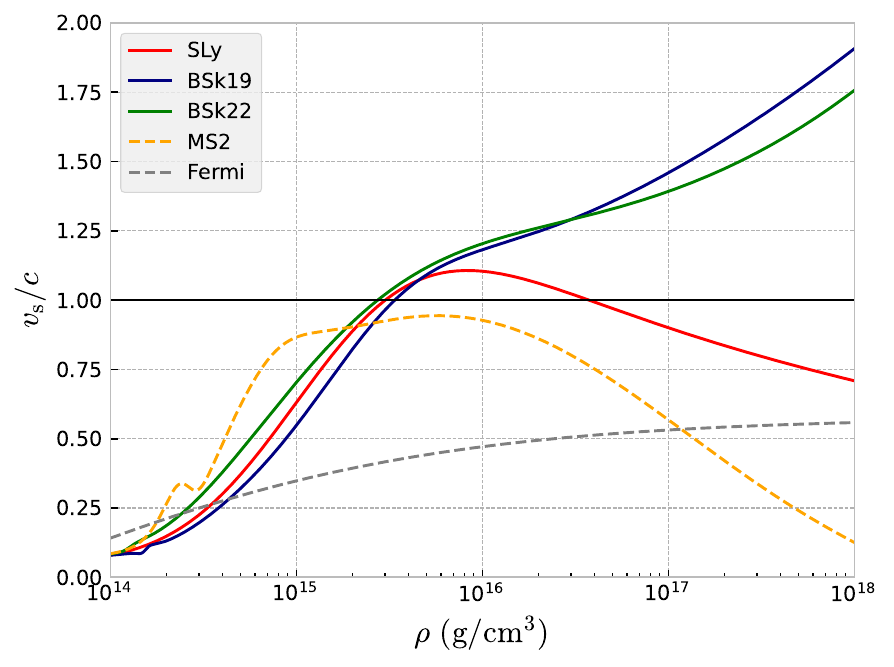}
    \caption{The speed of sound $v_{\rm s}$ as a function of the density $\rho$ for the EOSs SLy, BSk19, BSk22, and MS2 equations of state. The horizontal line marks the causal limit $v_{\rm s} = c$.}
    \label{fig:sound-speed}
\end{figure}

An illustration for this expression is found in Fig. \ref{fig:sound-speed}. One can see that for SLy, BSk19, and BSk22 there exists a critical  density above which the speed of sound will be greater than the speed of light (violating causality). This is due to the nonrelativistic formalism behind these EOSs (in contrast to the MS2 EOS, for which this problem does not occur). The causal limit puts constraints on the density for the EOSs considered, with the maximum densities allowed being $3.007$, $3.381$, and $2.737$ (in units of $10^{15} \ {\rm g/cm^3}$) for the EOSs SLy, BSk19, and BSk22, respectively. Finally, the speed of sound for the SLy EOS is again less than the speed of light at $\rho = 3.747 \times 10^{16} \ {\rm g/cm^3}$.

\subsection{Numerical solutions}

We solve numerically the system of equations \eqref{eq:df-static}, \eqref{eq:dchi-static}, and \eqref{eq:dP-static}  using the fourth-order Runge Kutta method for a given value of central density. For each EOS, we choose a central density $\rho_{\rm c}$, compute the corresponding central pressure $P_{\rm c}$ using the EOS, and integrate with the initial conditions: $f(0) = 1$, $\chi_{\rm num}(0) = 0$, and $P(0) = P_{\rm c}$, integrating outward until the pressure vanishes at some radius $R$, i.e., $P(R)=0$, which defines the surface of the star. The mass of the star is obtained by matching the interior and exterior solutions, solving for $M$ in the following equation:
\begin{equation}
    f_{\rm num}(R) = 1 + \frac{R^2}{2\alpha} \left( 1 - \sqrt{1 + \frac{8\alpha GM}{c^2 R^3}}\right),
\end{equation}
with $f_{\rm num}(R)$ the numerical value of $f$ at the surface of the star. We proceed similarly for values of central density between $\rho_{\rm c} = 2 \times 10^{14} \, {\rm g/cm^3}$ and $\rho_{\rm c} = 1 \times 10^{19} \, {\rm g/cm^3}$. Not all densities in this interval are allowed in all cases, however. For $\alpha < 0$ there is, according to \eqref{eq:frexp}, a maximum critical value $\rho_{\rm c}$ such that  $1 + (32 \pi G \alpha \rho_{\rm c})/(3c^2) \leq 0$. On the other hand, for $\alpha > 0$, we increase $\rho_c$ up to the point when the function $f$ approaches zero, i.e., the solution approaches a black hole.

\begin{figure}
\centering
\subfigure[SLy]{\includegraphics[width=0.45\linewidth]{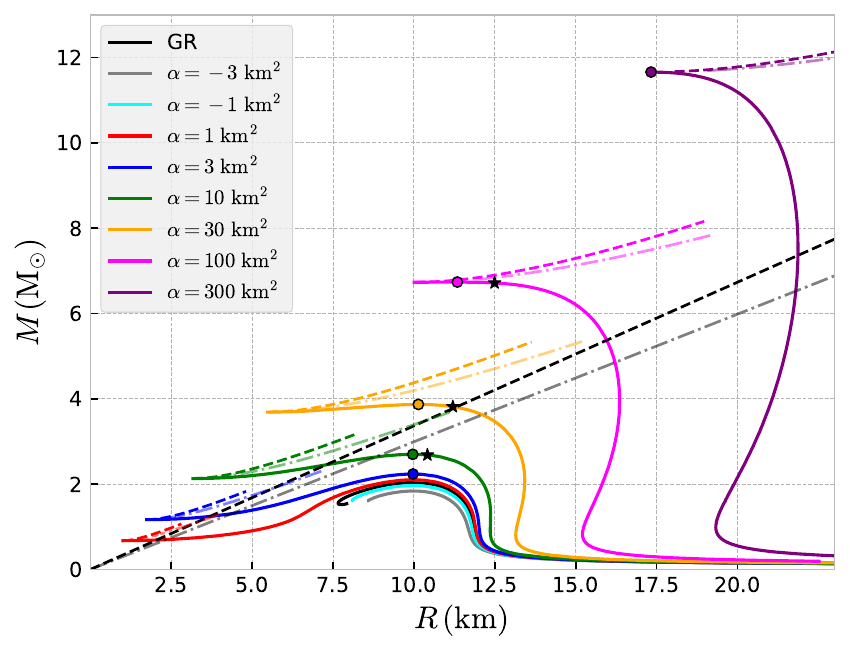}
\label{fig:MR-SLy}}
\subfigure[SLy]{\includegraphics[width=0.45\linewidth]{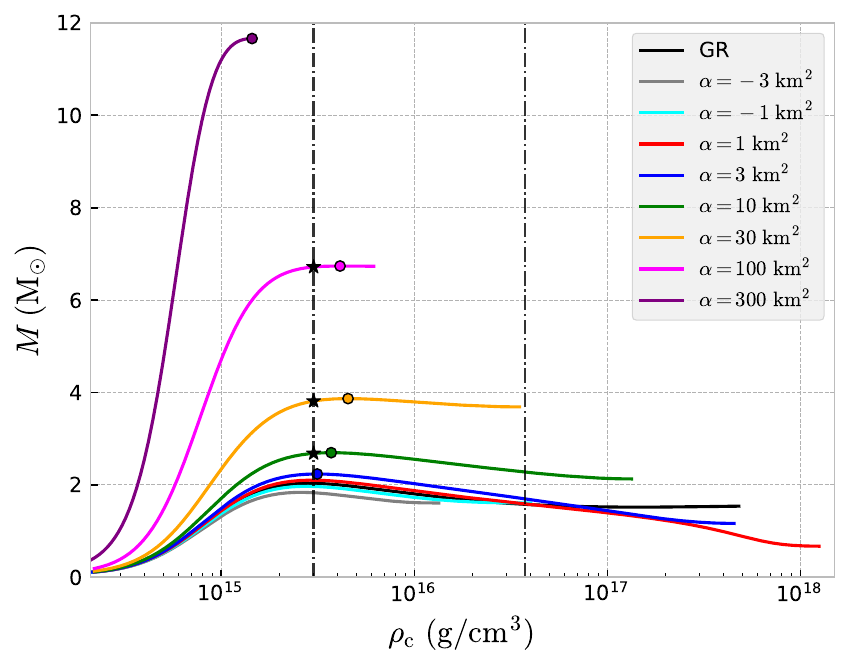}
\label{fig:Mrho-SLy}}
\subfigure[BSk19]{\includegraphics[width=0.45\linewidth]{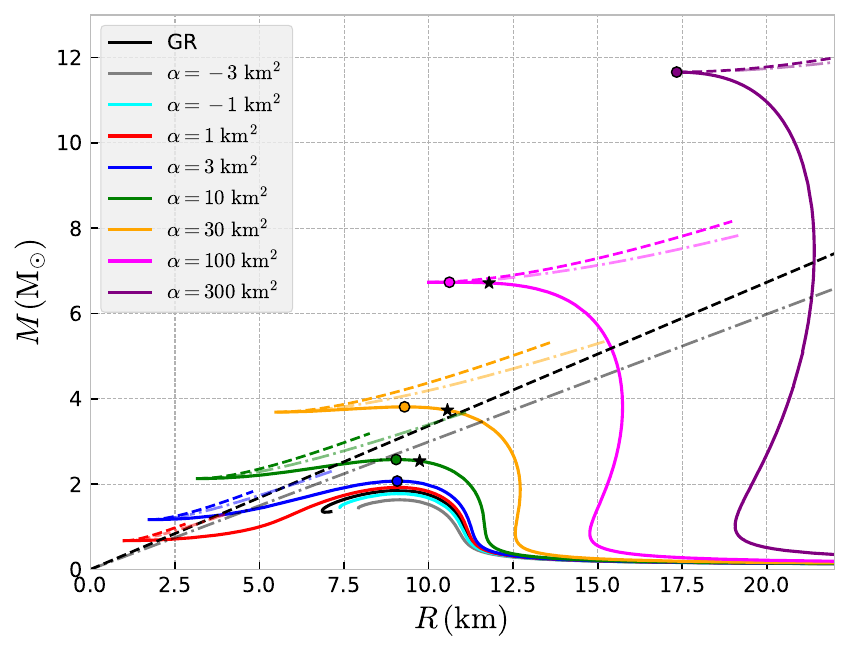}
\label{fig:MR-BSk19}}
\subfigure[BSk19]{\includegraphics[width=0.45\linewidth]{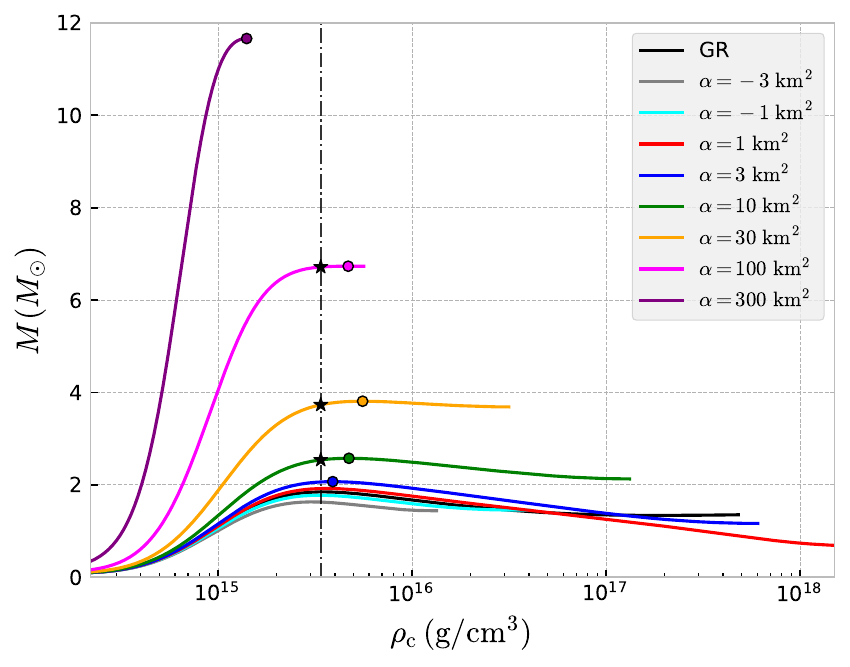}
\label{fig:Mrho-BSk19}}
\subfigure[BSk22]{\includegraphics[width=0.45\linewidth]{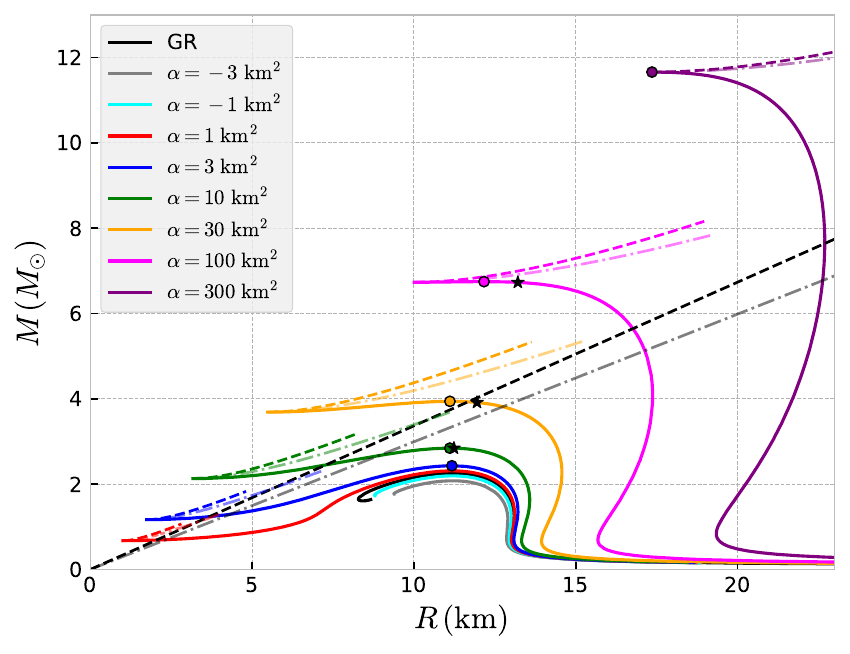}
\label{fig:MR-BSk22}}
\subfigure[BSk22]{\includegraphics[width=0.45\linewidth]{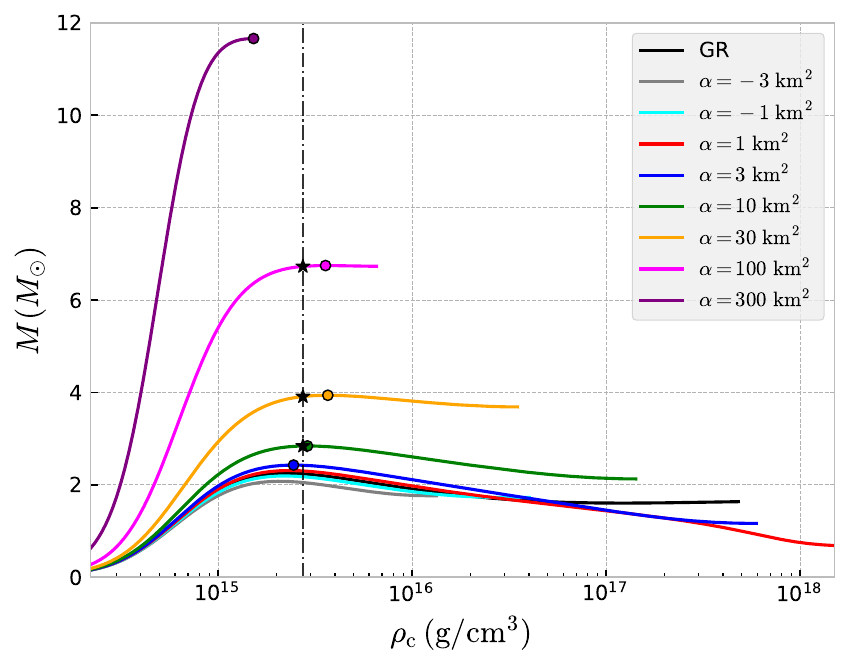}
\label{fig:Mrho-BSk22}}

\caption{Mass versus radius and central density curves for neutron stars using the SLy, BSK19, and BSk22 EOSs in GR (black solid line) and in 4DEGB gravity for different values of $\alpha$ (colorful solid lines). The starred points mark the NS configuration where the speed of sound is equal to the speed of light. In plots (a), (c), and (e), the dashed lines represent the mass-radius curves for the relevant black holes, and the dash-dotted lines correspond to the Buchdahl limits in these two theories of gravity. 
    In plots (b), (d), and (f), the vertical lines mark the central density where the speed of sound is equal to the speed of light.}
\label{fig:MvsR-SLy-BSk}
\end{figure}

\begin{figure}
\centering
\subfigure[MS2]{\includegraphics[width=0.45\linewidth]{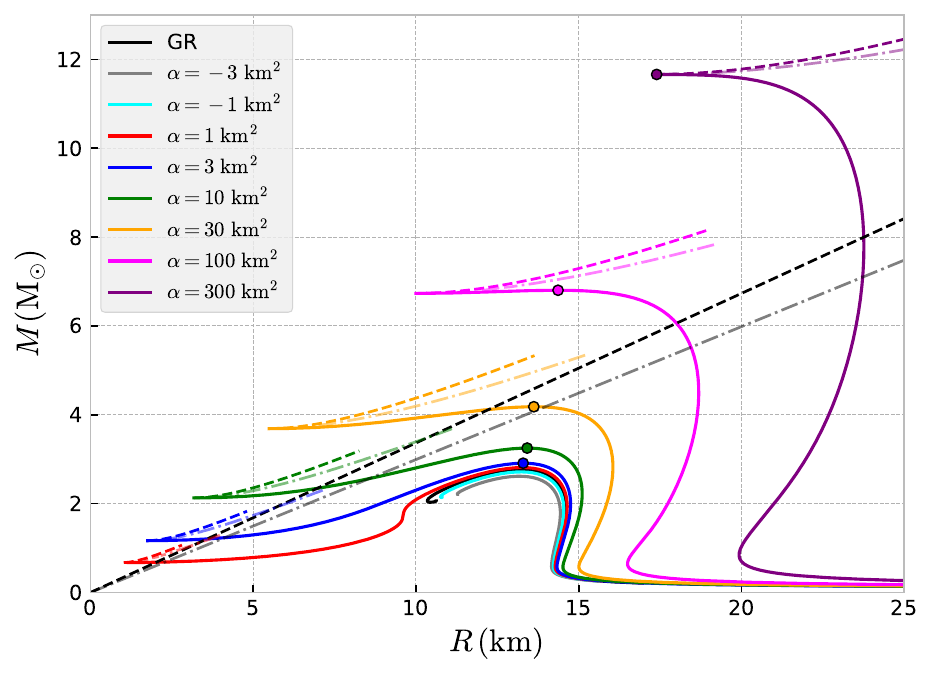}
\label{fig:MR-MS2}}
\subfigure[MS2]{\includegraphics[width=0.45\linewidth]{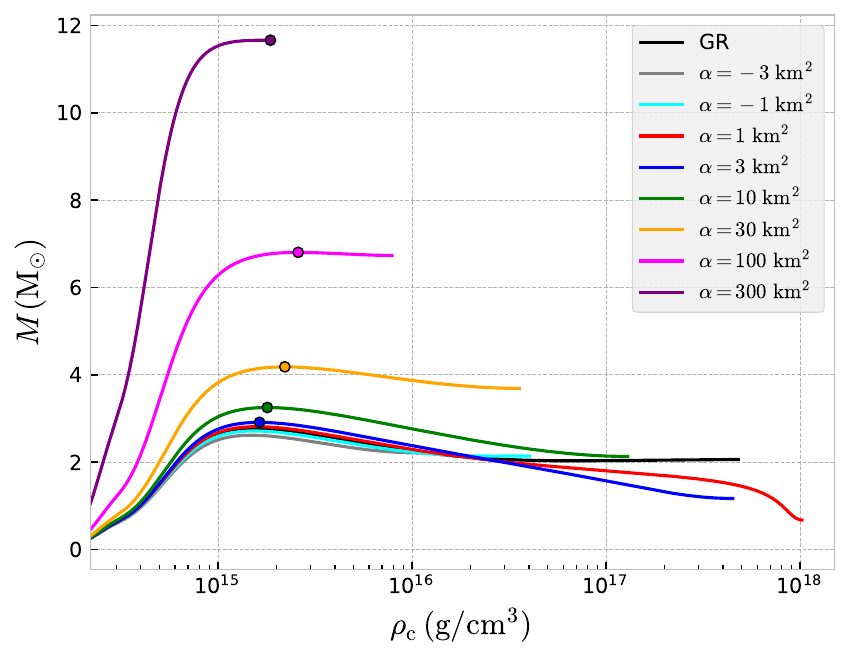}
\label{fig:Mrho-MS2}}
 
\caption{Mass versus radius and central density curves for neutron stars using the MS2 EOS in GR (black solid line) and in 4DEGB gravity for different values of $\alpha$ (colorful solid lines). The starred points mark the NS configuration where the speed of sound is equal to the speed of light. In plot (a) the dashed lines represent the mass-radius curves for the relevant black holes, and the dash-dotted lines correspond to the Buchdahl limits in these two theories of gravity.  In plot (b) we note the lack of a vertical line marking the transition from subluminal to superluminal sound speeds as this EOS always respects causality.}
\label{fig:MvsR-MS2}
\end{figure}

In Fig. \ref{fig:MR-SLy}, we show the mass-radius relation of stars for the SLy EOS, reproducing the same result reported in Ref. \cite{Charmousis2022}, and in Figs. \ref{fig:MR-BSk19} and \ref{fig:MR-BSk22} for BSk19-22 EOSs, using different values of the coupling constant $\alpha$. First, we see that positive values of $\alpha$ increase the mass of the NS for any given value of radius with respect to GR commensurate with positive $\alpha$ weakening gravity. The BSk19 EOS predicts smaller and lighter neutron stars, while BSk22 predicts larger, more massive neutron stars (see Fig. \ref{fig:MvsR-SLy-BSk} for a clear comparison using $\alpha = 10 \ {\rm km^2}$). This is due to the stiffness of the EOS increasing from BSk19 to BSk22. Therefore, even though the SLy EOS provides a unified model of matter, the BSk models are more flexible and offer a broader range of possible neutron star configurations due to their different versions. Finally, in Fig. $\ref{fig:MR-MS2}$ the same results are shown for the MS2 EOS, which are strikingly similar to those shown for the nonrelativistic equations of state.  We note that, when the large $\alpha$ solutions have a maximum mass near the black hole horizon with the nonrelativistic EOSs, this maximum mass point is relatively further from this intersection in MS2. 

\begin{figure}
\centering
\subfigure[]{\includegraphics[width=0.45\linewidth]{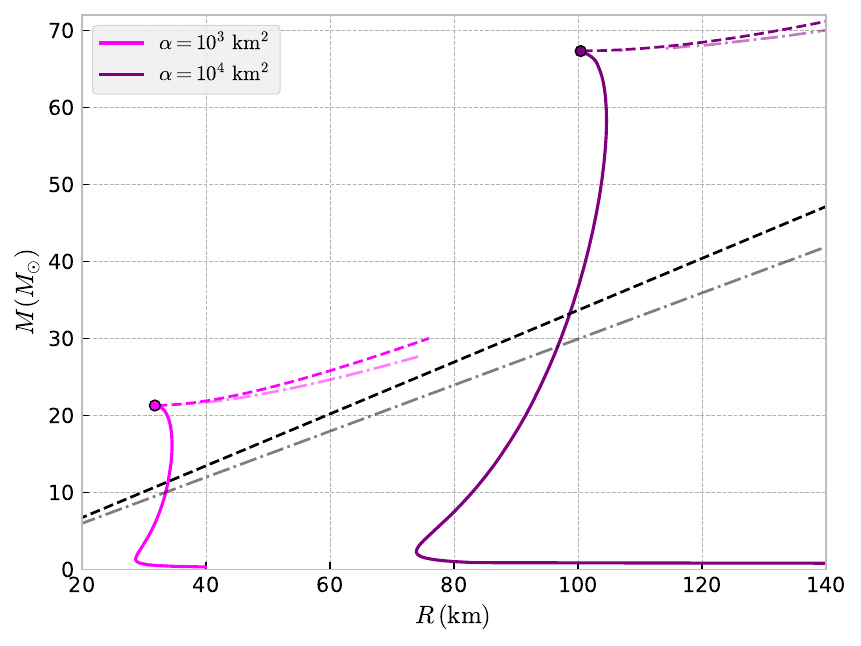}
\label{fig:MR-SLy-large-alpha}}
\subfigure[]{\includegraphics[width=0.45\linewidth]{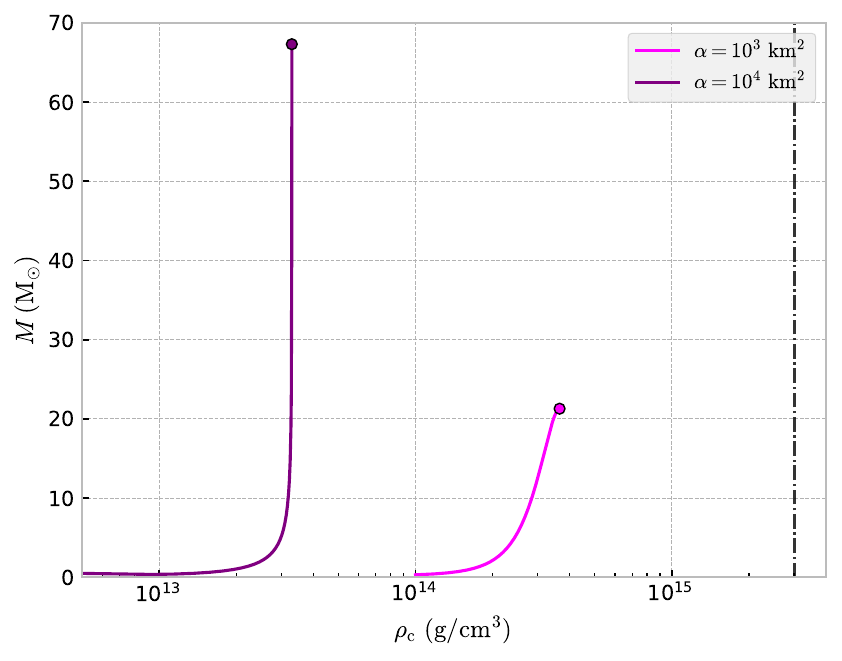}
\label{fig:Mrho-SLy-large-alpha}}

\caption{Mass versus radius and central density curves for NS using the SLy EOS in 4DEGB gravity for $\alpha = 10^{3} \ {\rm km^2}$ and $\alpha = 10^{4} \ {\rm km^2}$. In plot (a) the dashed lines represent the black hole horizons, and the dash-dotted lines correspond to the Buchdahl limits in these two theories. The maximum mass solutions effectively overlap with the minimum mass black hole horizon/Buchdahl bound intersection point. These solutions contain a wide range of ECOs, which are disallowed in standard GR. In plot (b) the vertical line marks the central density where the speed of sound is equal to the speed of light. We notice that for this larger coupling to the higher curvature gravity the breaking of causality stops being a problem as the relevant central densities are smaller.}
\label{fig:MvsR-SLy-large-alpha}
\end{figure}

In Figs. \ref{fig:Mrho-SLy}, \ref{fig:Mrho-BSk19}, and \ref{fig:Mrho-BSk22}, the mass versus central density curves are plotted for the SLy, BSk19, and BSk22 equations of state. The vertical dashed lines demarcate the central density where the speed of sound is equal to the speed of light. In Fig. \ref{fig:Mrho-SLy} the solutions between the two vertical lines do not satisfy the condition of maximum speed of sound. Similarly, in Figs. \ref{fig:Mrho-BSk19} and \ref{fig:Mrho-BSk22} the solutions lying to the right of the vertical line are also excluded. This condition defines maximum masses for which the solution is reliable. For example, in Fig. \ref{fig:Mrho-SLy}, the maximum mass point is located to the right of the first vertical line for $\alpha \gtrsim 3 \ {\rm km^2}$, but to the left for smaller $\alpha$ (for instance, $\alpha = 1 \ {\rm km^2}$). In other words, as we increase the value of the coupling $\alpha$ the maximum mass points of the $M-\rho_{\rm c}$ curves move to the left, toward the vertical causality line. Therefore, there must be a value of $\alpha$ such that all neutron star solutions satisfy the causality condition, which is the case for $\alpha = 300 \ {\rm km^2}$, and for $\alpha = 10^3, 10^4 \ {\rm km^2}$, see Fig. \ref{fig:MvsR-SLy-large-alpha} . Finally in Fig. $\ref{fig:Mrho-MS2}$ the same data is plotted for the relativistic MS2 equation of state - in this case the causality line is absent as this EOS always respects causality.

\begin{figure}
    \centering
    \includegraphics[width=0.75\linewidth]{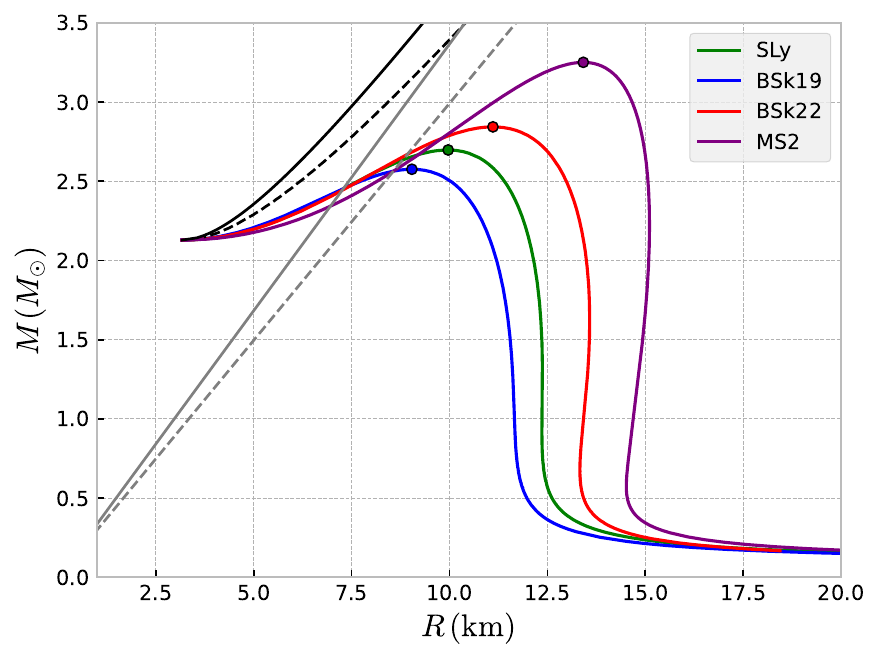}
    \caption{Mass-radius relation for neutron stars using the SLy, BSk19-22 and MS2 EOSs in 4DEGB gravity for $\alpha = 10 \ {\rm km^2}$. The black solid line corresponds to the mass-radius relation for the black holes, and the dashed line represents the 4DEGB Buchdahl limit for that particular value of $\alpha$. In contrast, the gray solid and dashed lines show the mass-radius relation for the black holes and the Buchdahl limit, respectively, in general relativity. The colored dots mark the maximal mass point of a given curve.}
    \label{fig:MvsR-SLy-BSk}
\end{figure}

Charmousis, Lehébel, Smyrniotis and Stergioulas \cite{Charmousis2022} pointed out that at high central densities, the neutron star equilibrium configurations approach the black hole limit asymptotically, and we have numerically confirmed that these two sequences (NS and BH configurations) become arbitrarily close near the minimum mass black hole of the theory. This result is compatible with the modified Buchdahl bound of this theory given by \cite{CHAKRABORTY2020}
\begin{equation}
    \sqrt{1 - \mu R^2} (1 + \alpha \mu) > \frac{1}{3}(1 - \alpha \mu),
\end{equation}
where 
\begin{equation}
    \mu := \frac{1}{2\alpha} \left(\sqrt{1 + \frac{8\alpha GM}{c^2 R^3}} - 1\right)
\end{equation}
for a star of radius $R$ and mass $M$. This compatibility is due to the fact that the Buchdahl bound intersects the black hole horizon \eqref{eq:BH-horizon} at the minimum black hole mass allowed by this theory, $M_{\rm int} = c^2 \sqrt{\alpha}/G$, which was first discussed in Ref. \cite{Gammon2023}. These results imply that it is possible to have compact objects in 4DEGB whose radii are smaller than those of the GR Buchdahl bound $R \geq 9GM/4c^2$ or even that of the Schwarzschild radius $r_{\rm s} = 2GM/c^2$. 

\section{Adiabatic radial oscillations}\label{sec:radosc}

Equilibrium configurations of a compact star are obtained by integrating Eqs. \eqref{eq:df-static}-\eqref{eq:dP-static}. In what follows, we will study whether that equilibrium is stable under adiabatic radial oscillations following the approach by Chandrasekhar \cite{Chandrasekhar1964}. In GR, a necessary but insufficient condition for the stability of compact stars is $dM/d\rho_{\rm c} < 0$ \cite{Glendenning}, corresponding to the parts of the solution curves before the maximum mass point. There is no  similar theorem in 4DEGB, so we will investigate if this holds in spite of the modifications to gravity.

Consider a perfect fluid sphere oscillating radially with a small amplitude. Since the oscillations are radial, the spacetime preserves its spherical symmetry. Thus, we can express the line element as follows:
\begin{equation}
\label{eq:perturbed-metric}
ds^2 = - e^{\chi} f (dx^0)^2 + \frac{dr^2}{f} + r^2(d\theta^2 + \sin^2\theta d\varphi^2), 
\end{equation}
where $x^{0} = ct$, $\chi = \chi(t,r)$, and $f = f(t,r)$ are the new metric functions, which have now been perturbed. With this we can write
\begin{gather}
    \chi(t,r) = \chi_0(r) + \delta \chi(t,r), \\
    f(t,r) = f_0(r) + \delta f(t,r),
\end{gather}
where we have used the subscript $0$ to denote the field in the equilibrium configuration (nonperturbed). Similar expressions are constructed for the scalar field, the pressure, and the energy density. 

In addition, it is necessary to introduce another perturbation describing the oscillations: the radial displacement $\delta r$ of the fluid from its equilibrium position. 
Thus, a fluid element located at the radial coordinate $r$ in the equilibrium configuration is displaced to the new radial coordinate  $r + \delta r(t,r)$ at coordinate time $t$ in the perturbed configuration. Similarly, in the context of fluid mechanics, we define the \textit{Eulerian perturbations}, denoted by $\delta P$ and $\delta \epsilon$, as changes measured by a fixed observer at a point described by $(t,r,\theta,\varphi)$. Alternatively,  the \textit{Lagrangian perturbations}, denoted by $\Delta P$ and $\Delta \epsilon$,  are changes measured by an observer located at $r$ in the equilibrium configuration, but located at $r + \delta r(t, r)$ in the perturbed configuration. Eulerian and Lagrangian perturbations are related by
\begin{equation}
\Delta P(t,r) = P[t,r +\delta r(t,r)] - P_0(r) = \delta P + P_0' \delta r.   \label{eq:P-lagrangian}
\end{equation}

The Eulerian perturbations of the pressure and the energy density are given by \cite{MTW}
\begin{gather}
\delta P = \Gamma P_0 \left[ \frac{\delta f}{2 f_0} - \sqrt{f_0} r^{-2} \frac{\partial}{\partial r} \left( \frac{r^2 \delta r}{\sqrt{f_0}}\right) \right] - P_0' \delta r, \label{eq:EGB-delta-P}\\
\delta \epsilon = (\epsilon_0 + P_0) \left[ \frac{\delta f}{2 f_0} - \sqrt{f_0} r^{-2} \frac{\partial}{\partial r} \left( \frac{r^2 \delta r}{\sqrt{f_0}}\right) \right] - \epsilon_0' \delta r. \label{eq:EGB-delta-density}
\end{gather}

Solving  the component $(t r)$ of the field equations to first order, we find that 
\begin{equation}
    \delta f = \frac{8\pi G}{c^4} \frac{(\epsilon_0 + P_0)r^3 \delta r}{r^2 - 2\alpha f_0 + 2\alpha} = f_0 \chi_0' \delta r. \label{eq:EGB-delta-f}
\end{equation}

Equation \eqref{eq:EGB-delta-f} allows us to write the evolution of the perturbations $\delta P$ and $\delta \epsilon$ in terms of the fields in the equilibrium configuration and the radial displacement $\delta r$:
\begin{align}
    \delta P &= -\Gamma P_0 r^{-2} \sqrt{f_0} e^{\chi_0/2}  \sigma'  - P_0' \delta r,\label{eq:EGB-delta-P-2}  \\
    \delta \epsilon &= -(\epsilon_0 + P_0)  r^{-2} \sqrt{f_0} e^{\chi_0/2}  \sigma'  - \epsilon_0' \delta r, \label{eq:EGB-delta-densidad-2} 
\end{align}
where we have defined the \textit{normalized displacement function} \cite{MTW} as
\begin{equation}
    \sigma := \frac{r^2e^{-\chi_0/2}}{\sqrt{f_0}}  \delta r. \label{eq:zeta-def}
\end{equation}

The $(rr)$ component of \eqref{eq:metric-FEq} gives us $\partial \delta \chi/\partial r$ in terms of the $\delta \phi$ and $\delta r$. The equations corresponding to the other components are fully satisfied.

From the conservation of the energy-momentum tensor and the equation for the scalar field \eqref{eq:scalar-FEq}, properly linearized, we obtain the dynamic equations for $\sigma$ and $\delta \phi$, respectively,
\begin{gather}
 A_1(r) \frac{\partial^2 \sigma}{\partial t^2} + A_2(r) \frac{\partial^2 \sigma}{\partial r^2} + A_3(r) \frac{\partial \sigma}{\partial r} + A_4(r) \sigma + A_5(r) \frac{\partial \delta \phi}{\partial r} = 0, \label{eq:EGB-zeta} \\
B_1(r) \frac{\partial^2 \delta \phi}{\partial r^2} + B_2(r) \frac{\partial \delta \phi}{\partial r} + B_3(r) \frac{\partial \sigma}{\partial r} + B_4(r) \sigma = 0,    \label{eq:EGB-dphi}
\end{gather}
where the coefficients can be found in Appendix \ref{app:RO}.
 
\subsection{Boundary conditions}

Not all solutions of \eqref{eq:EGB-zeta} and  \eqref{eq:EGB-dphi}  are physically acceptable. In order to be realistic, the displacement function $\delta r$ must produce finite density and pressure perturbations at the center of the stars, which means 
\begin{equation}
    \lim_{r\to 0^{+}} \frac{\delta r}{r} <\infty. \label{eq:EGB-BC-xi}
\end{equation}

In terms of the function $\sigma$, this boundary condition translates to
\begin{equation}
    \lim_{r\to 0^{+}} \frac{\sigma}{r^3} <\infty. \label{eq:EGB-BC-zeta}
\end{equation}

In addition, the displacement function must also leave the pressure equal to zero at the surface of the star. Thus, from Eqs. \eqref{eq:P-lagrangian} and \eqref{eq:EGB-delta-P-2}, we must have
\begin{equation}
    \lim_{r \to R} \Delta P = \lim_{r \to R} \left[-\Gamma P_0 r^{-2} \sqrt{f_0} e^{\chi_0/2}  \sigma' \right] = 0. \label{eq:EGB-BC-pressure}
\end{equation}

\subsection{Numerical analysis}

Assuming a harmonic time dependence such as 
\begin{gather}
    \sigma(t,r) = u(r) e^{-i\omega t}, \\
    \delta \phi(t,r) = \varphi(r) e^{-i\omega t},
\end{gather}
the evolution equations \eqref{eq:EGB-zeta} and \eqref{eq:EGB-dphi} reduce to
\begin{gather}
       A_2 u'' + A_3 u' + [A_4 - \omega^2 A_1] u + A_5 \varphi' = 0, \label{eq:EGB-zeta-2.0}\\
   B_1 \varphi'' + B_2 \varphi' + B_3 u' + B_4 u = 0.\label{eq:EGB-dphi-2.0}
\end{gather}

Note that Eq. \eqref{eq:EGB-zeta-2.0} can be written as
\begin{equation}
    -\frac{d}{dr}\left(A_2 u'\right) =   [A_4 - \omega^2 A_1] u + A_5 \varphi'. \label{eq:EGB-zeta-3.0}
\end{equation}

Multiplying \eqref{eq:EGB-dphi-2.0} by the integrating factor,
\begin{equation}
    \eta(r) = \exp\left(\int \frac{B_2}{B_1} dr\right) = \sqrt{f_0}\left(r f_0 \chi_0' + r f_0' + 2(\sqrt{f_0} - f_0) \right) e^{\chi_0/2},
\end{equation}
 and then dividing it by $B_1$, we obtain
\begin{equation}
    \exp\left(\int \frac{B_2}{B_1} dr\right) \varphi'' + \frac{B_2}{B_1} \exp\left(\int \frac{B_2}{B_1} dr\right)\varphi' + \frac{B_3}{B_1} \eta u' + 
 \frac{B_4}{B_1} \eta u = 0. \label{eq:EGB-dphi-3.0}
\end{equation}

We can write Eq. \eqref{eq:EGB-dphi-3.0} as follows:
\begin{equation}
    \frac{d}{dr} \left(  \eta \varphi' \right) + \frac{B_3}{B_1} \eta u' + \frac{B_4}{B_1} \eta u = 0. \label{eq:EGB-dphi-3.0}
\end{equation}

Defining $v := - A_2 u'$ and $\psi = \eta \varphi'$, the system of second order differential equations given by \eqref{eq:EGB-zeta-3.0} and \eqref{eq:EGB-dphi-3.0} can be written as the following system of first order differential equations for the functions $u$, $v$ and $\psi$:
\begin{gather}
    u' = c_1 v, \label{eq:EGB-EDO-u}\\
    v' = [c_2 + \omega^2 c_3] u + c_4 \psi,  \label{eq:EGB-EDO-v}\\
    \psi' = c_5 v + c_6 u,  \label{eq:EGB-EDO-psi}
\end{gather}
where the coefficients are given by
\begin{gather}
    c_1 = -\frac{1}{A_2}, \\
    c_2 = A_4, \\
    c_3 = - A_1, \\
    c_4 = \frac{A_5}{\eta} = \frac{2\alpha(\epsilon_0 + P_0)e^{\chi_0/2}}{f_0(r^2 - 2\alpha f_0 + 2\alpha)}, \\
    c_5 = \frac{B_3}{B_1A_2} \eta =   \frac{\left(r f_0 \chi_0' + r f_0' + 2(\sqrt{f_0} - f_0) \right) \chi_0' }{2\Gamma P_0 \sqrt{f_0} e^{\chi_0/2}r}, \\
    c_6 = - \frac{B_4}{B_1} \eta = -\frac{ B_4 e^{\chi_0/2}}{8 r^4 \sqrt{f_0}}.
\end{gather}

Near the origin we expand the functions $u$, $v$, $\psi$, $\Gamma$, $P_0$, $\epsilon_0$, $f_0$ and $\delta r$ as per Eq. \eqref{eq:Taylor-expansion}. Using the boundary condition \eqref{eq:EGB-BC-zeta}, we have $u_0 = u_1 = u_2 = 0$. In addition, we know from the nonperturbed case that $f_0(0) = 1$, $f_0'(0) = 0$, and $\chi_0'(0) = 0$. Therefore,
\begin{gather}
    u(r) = u_3 r^3 + \mathcal{O}(r^4), \\
    f_0(r) = 1 + f_{0,2} r^2 + f_{0,3} r^3 + \mathcal{O}(r^4), \\
    \chi_0(r) = \chi_{\rm c} + \chi_{0,2} r^2 + \chi_{0,3} r^3 + \mathcal{O}(r^4).
\end{gather}

Replacing these expansions in the Eqs. \eqref{eq:EGB-EDO-u}-\eqref{eq:EGB-EDO-psi}, we obtain 
\begin{gather}
    v(r) = 3u_3 \Gamma(0) P_{\rm c} e^{3\chi_{\rm c}/2} + \mathcal{O}(r), \\
    \psi(r) = \psi_2 r^2 + \mathcal{O}(r^3).
\end{gather}

We can choose the initial values as $u(0) = 0$, $v(0) = 1$, and $\psi(0) = 0$ without loss of generality because the differential equations \eqref{eq:EGB-EDO-u}-\eqref{eq:EGB-EDO-psi} are linear; this is equivalent to defining new variables $\tilde{u} = u/(3\Gamma(0) P_{\rm c} e^{3\chi_{\rm c}/2})$ and $\tilde{v} = v/(3\Gamma(0) P_{\rm c} e^{3\chi_{\rm c}/2})$. At the surface of the star the condition \eqref{eq:EGB-BC-pressure} translates to $v(R) = 0$. Equation \eqref{eq:EGB-zeta-3.0} with these boundary conditions would be a Sturm-Liouville problem if it were not for the differential equation \eqref{eq:EGB-dphi-3.0} for the scalar field coupled to \eqref{eq:EGB-zeta-3.0}. 

\begin{figure}
    \centering
    \includegraphics[width=0.75\linewidth]{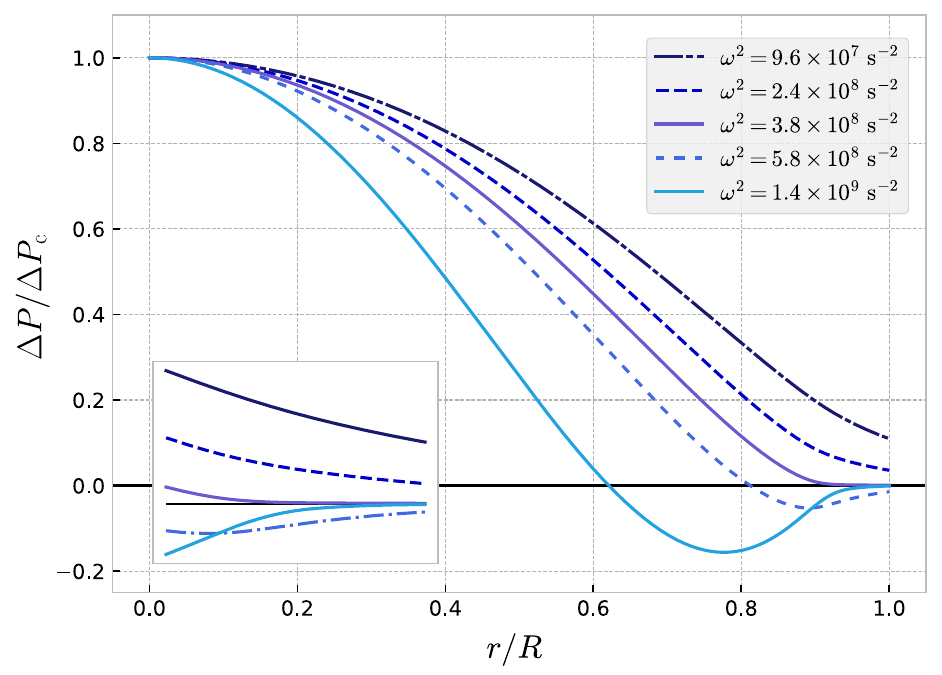}
    \caption{Normalized Lagrangian perturbation of the pressure inside the star for several test values of $\omega^2$ with central density $\rho_{\rm c} = 7.47 \times 10^{14} \ {\rm g/cm^3}$. The inset shows the behavior near $r/R=1$.}
    \label{fig:delta-P}
\end{figure}

To find the appropriate numerical method to integrate the system of Eqs. \eqref{eq:EGB-EDO-u}-\eqref{eq:EGB-EDO-psi}, we assume that there is an ordered set of frequencies $\omega_0^2 < \omega_{1}^2 < \omega_{2}^2 <\cdots$ such that the $n$-th frequency corresponds to an eigenfunction with $n$ nodes that satisfies the boundary conditions of the problem. We can observe this behavior in Fig. \ref{fig:delta-P}, which shows solutions for the Lagrangian perturbation of the pressure for some values of $\omega^2$. This property is typical of a Sturm-Liouville problem, and so we use a standard method to solve this problem: the shooting method. This method consists of starting with a trial value of $\omega^2$, integrating toward its surface, and searching for the value of $\omega_n^2$ such that the boundary condition at the surface $v(R) = 0$ is satisfied and the solution has $n$ nodes. 

Consequently, under radial oscillations, we say:
\begin{itemize}
    \item[(i)] The star is unstable if any of the eigenvalues $\omega_{n}^2$ are negative, since we would have purely imaginary frequencies and the perturbation of the star would grow exponentially with the amplitude of the oscillation as $e^{|\omega|t}$.

    \item[(ii)] The star is stable if all of the eigenvalues $\omega_{n}^2$ are positive, since the frequency is real and any perturbation of the star will oscillate as $e^{i\omega t}$.
\end{itemize}

Finally, one of the important consequences of the ordered set of frequencies is that if the fundamental radial mode of a star is stable ($\omega_0^2 > 0$), then all radial modes are stable. In contrast, if the star is radially unstable, the fastest growing instability will occur through the fundamental mode  ($\omega_0^2$ is ``more negative'' than the other $\omega_{n}^2$).

\begin{figure}
\centering
\subfigure[Radial displacement]{\includegraphics[width=0.31\linewidth]{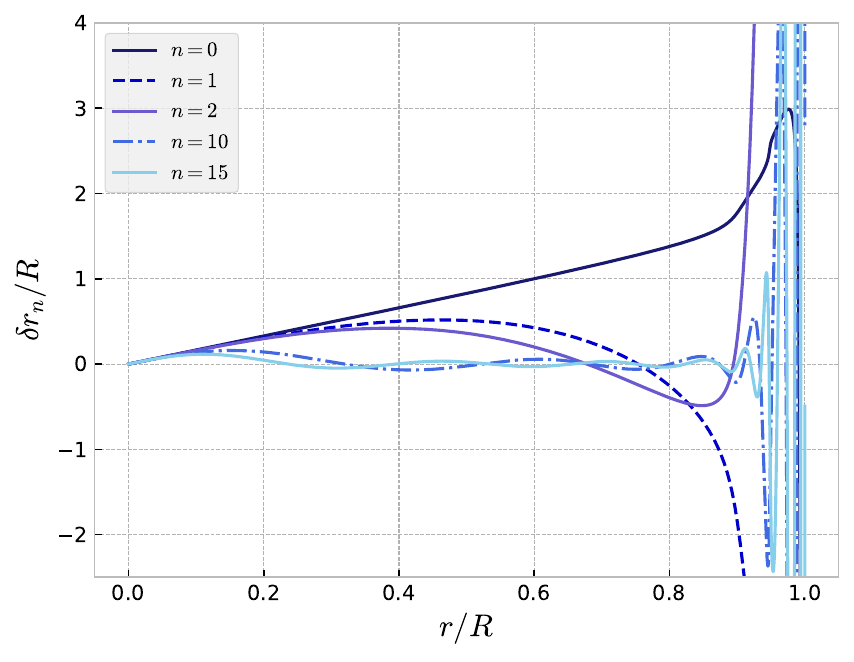}
\label{fig:delta-r-modes}}
\subfigure[Lagrangian perturbation of the pressure]{\includegraphics[width=0.31\linewidth]{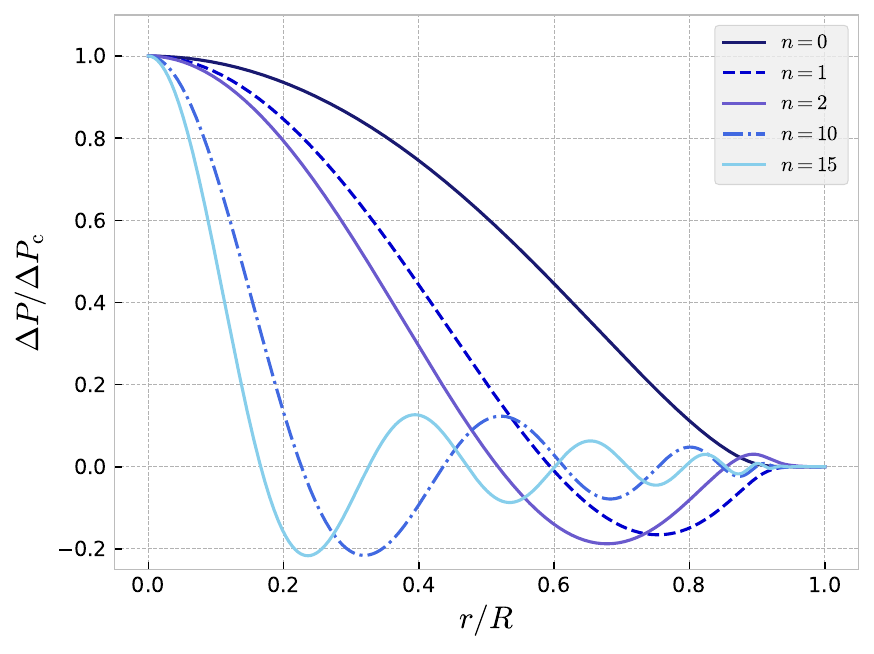}
\label{fig:delta-P-modes}}
\subfigure[Derivative of the perturbation of the  scalar field]{\includegraphics[width=0.31\linewidth]{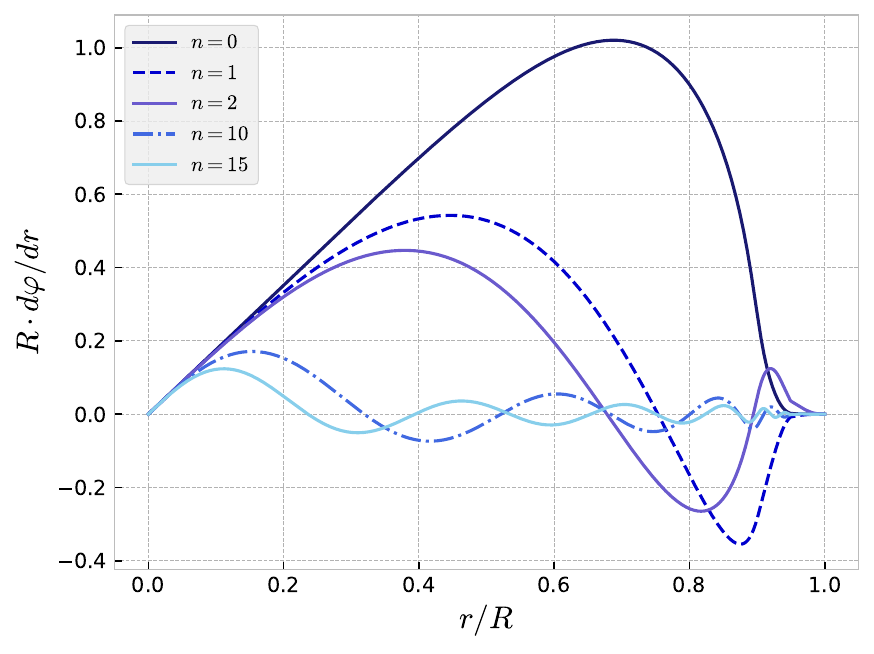}
\label{fig:delta-phi-modes}}

\caption{Plots of the radial displacement modes $\delta r_n$, the Lagrangian perturbation modes of the pressure $\Delta P_n$ and the derivative of the perturbation of the scalar field modes $d\varphi_n/dr$ with respect to the normalized radial coordinate $r/R$ for the fundamental mode ($n = 0$) and the $n$-overtones ($n = 1,2,10,15$) in a NS of mass $1.08 \ { M_{\odot}}$ and radius $R = 12 \ {\rm km}$ using the SLy EOS for $\alpha = 10 \ {\rm km^2}$.}
\label{fig:radial-profiles-modes}
\end{figure}

As a demonstration we have computed the radial profiles of the radial displacement modes $\delta r_n$, the Lagrangian perturbation modes of the pressure $\Delta P_n$, and the derivative of the perturbation of the scalar field modes $d\varphi_n/dr$ for the fundamental mode ($n = 0$) and the $n$-overtones ($n = 1,2,10,15$) in a NS of mass $1.08 \ { M_{\odot}}$ and radius $R = 12 \ {\rm km}$ using the SLy EOS for $\alpha = 10 \ {\rm km^2}$, shown in Fig. \ref{fig:radial-profiles-modes}. We observe that the amplitude of the radial displacement grows as the radial coordinate increases, whereas $\Delta P$ and $d\varphi/dr$ oscillate with a decaying amplitude before vanishing at the surface of the star. Nevertheless, all of the functions are smooth with respect to the radial coordinate. For higher-order modes, some of the nodes move across the core-crust transition and lie in the crust ($0.9 R \lesssim r \leq R$ \cite{Shapiro1983}), where the radial displacement changes sign rapidly with a large amplitude, but $\Delta P_n$ possesses a small amplitude in the crust. These results were observed in GR for the SLy EOS in Ref. \cite{Sen2023} and for a $f(R,T)$ gravity in Ref. \cite{Pretel2021}. The apparent divergence of the radial displacement at the surface of the star is due to numerical error and depends on the number of integration points and the tolerances used in the numerical calculation. This, however, does not affect the values of eigenfrequencies.

\begin{figure}
    \centering
    \includegraphics[width=0.75\linewidth]{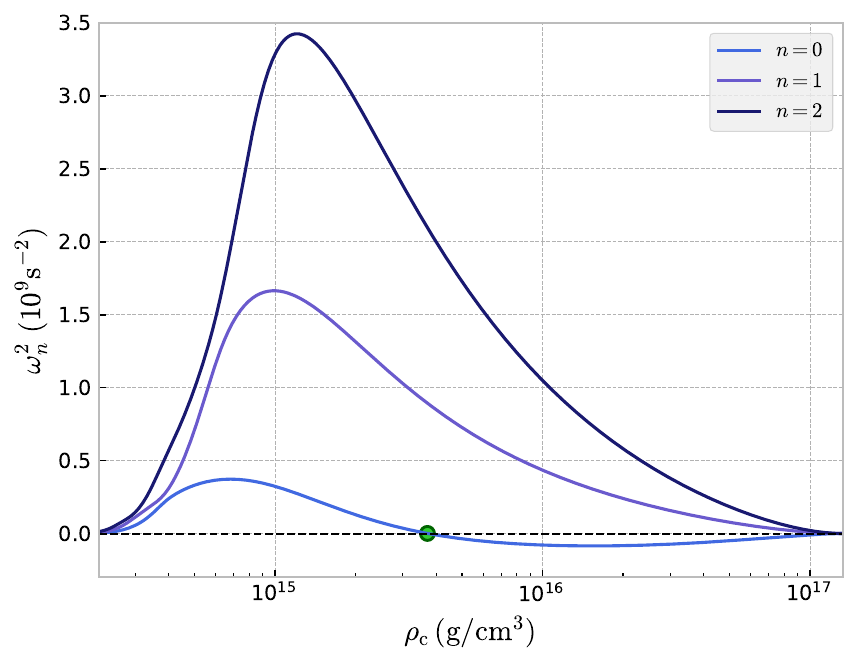}
    \caption{Squared frequencies of the first three oscillations modes for $\alpha = 10 \ {\rm km^2}$ using the SLy EOS. The green point indicates where the squared frequency of the fundamental mode becomes negative, which coincides with the maximum value of the mass $2.70 \ M_{\odot}$.}
    \label{fig:Modes-SLy}
\end{figure}

\begin{figure}
\centering

\subfigure[SLy]{\includegraphics[width=0.4\linewidth]{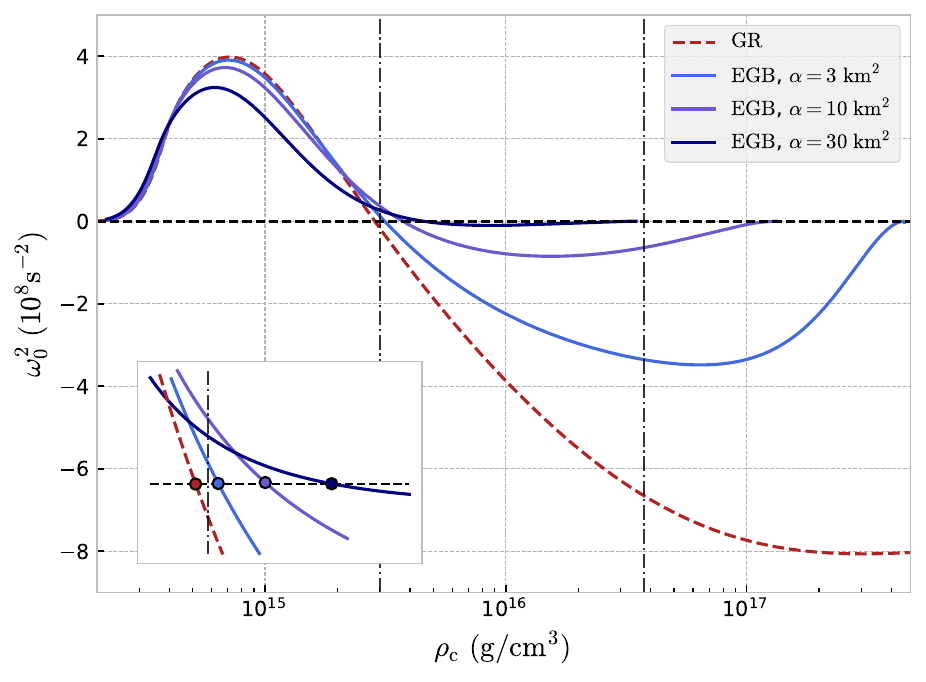}
\label{fig:Fundamental-Mode-SLy}}
\subfigure[SLy]{\includegraphics[width=0.4\linewidth]{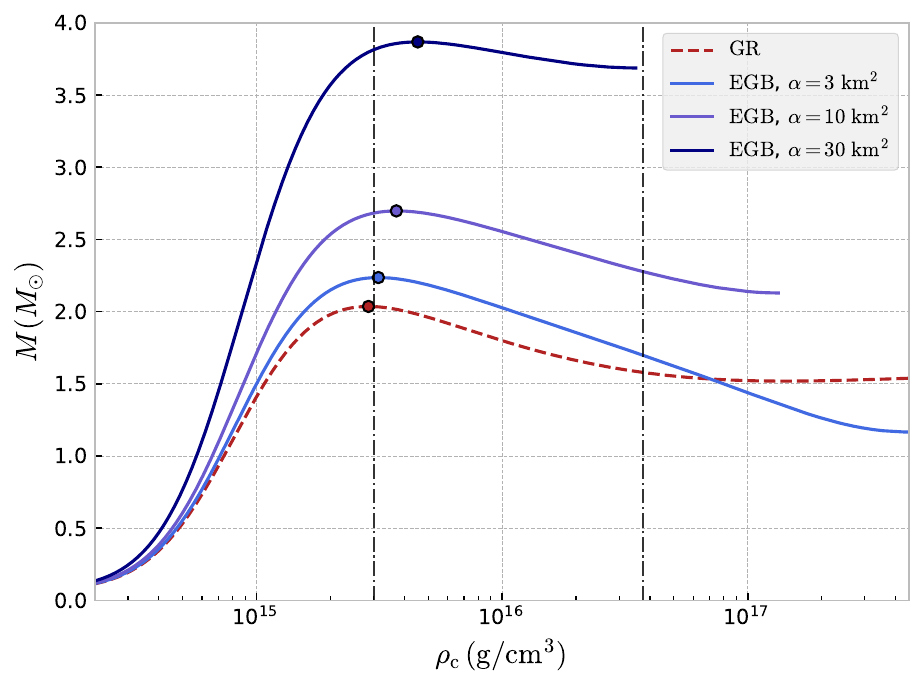}
\label{fig:M_rho-SLy}}

\subfigure[BSk19]{\includegraphics[width=0.4\linewidth]{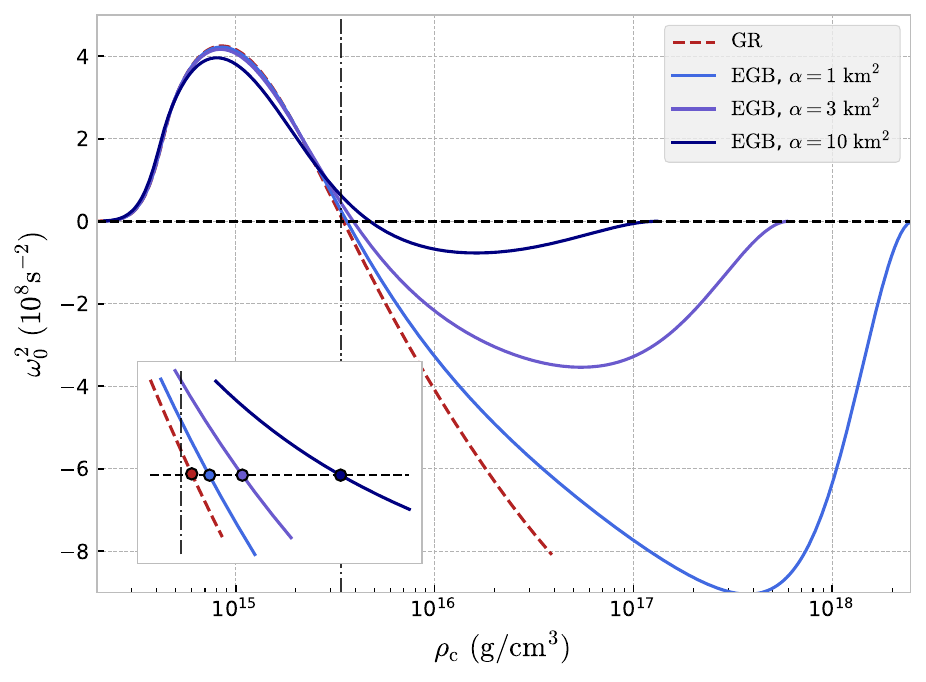}
\label{fig:Fundamental-Mode-BSk19}}
\subfigure[BSk19]{\includegraphics[width=0.4\linewidth]{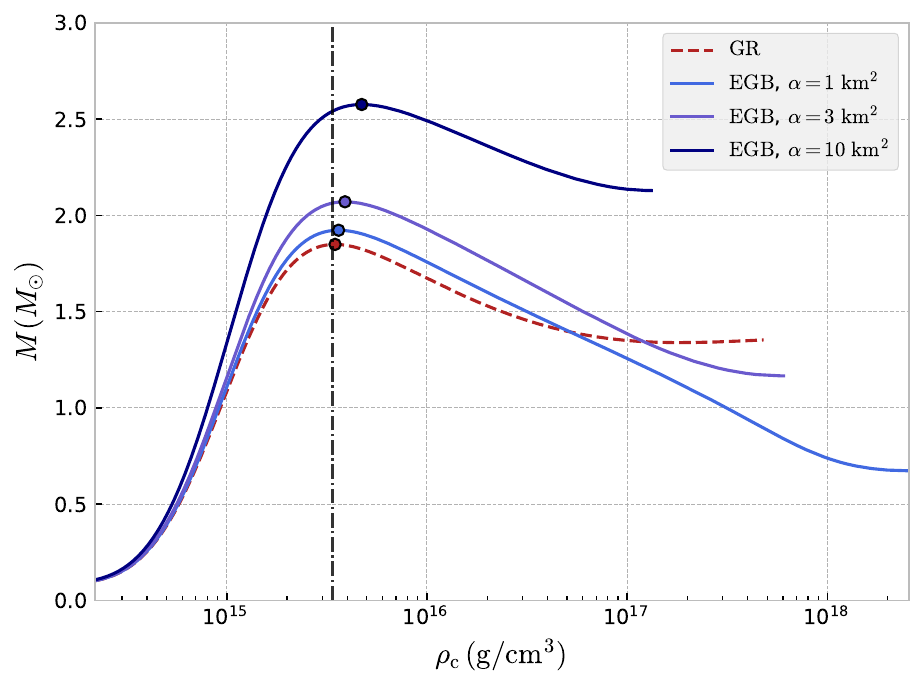}
\label{fig:M_rho-BSk19}}

\subfigure[BSk22]{\includegraphics[width=0.4\linewidth]{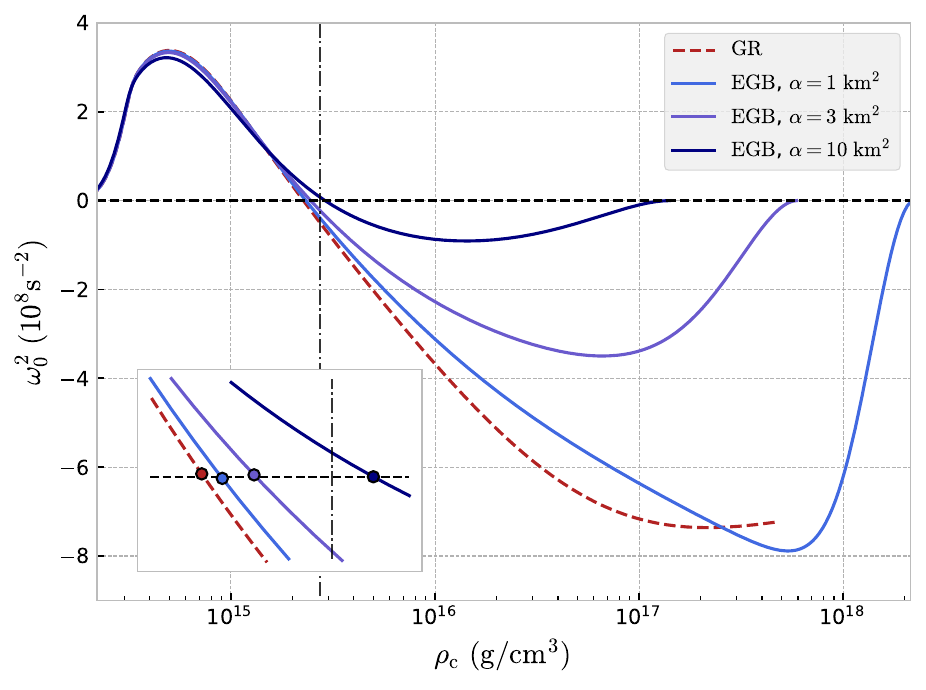}
\label{fig:Fundamental-Mode-BSk22}}
\subfigure[BSk22]{\includegraphics[width=0.4\linewidth]{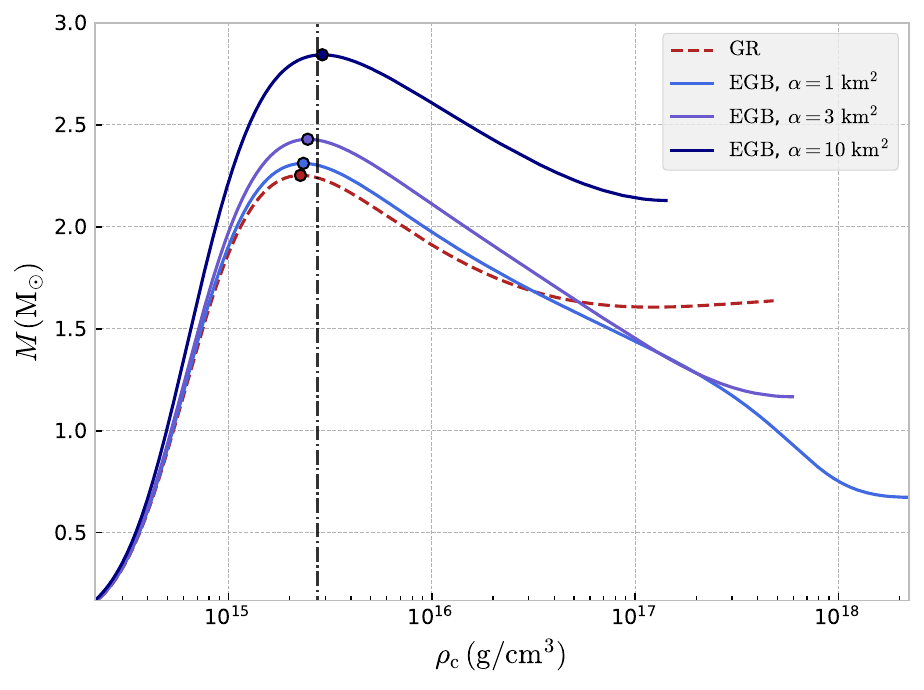}
\label{fig:M_rho-BSk22}}

\caption{Fundamental eigenfrequency and mass versus central density curves for neutron stars using the SLy, BSk19 and BSk22 EOSs in GR (red dashed line) and in 4DEGB for different values of $\alpha$ (blue lines). In plots (a), (c), and (e), the green circles mark the spot where the fundamental eigenfrequency is zero, while in plots (b), (d), and (f) they mark the maximum mass. The vertical lines mark the central density at which the speed of sound is equal to the speed of light. In plots (a) and (b) the second vertical line marks the central density at which the speed of sound is again subluminal.}
\label{fig:osc-modes}
\end{figure}

In Fig. \ref{fig:Modes-SLy} we show the eigenfrequencies of the first three oscillation modes as a function of central density for the SLy EOS with a value of $\alpha = 10 \ {\rm km^2}$. We observe that the squared frequencies of the overtones remain positive for all central densities, but for the fundamental mode there is a value of $\rho_{\rm c}$ for which $\omega_0^2$ becomes negative (and remains so). The point where $\omega_0$ is zero coincides with the maximum value of the mass, namely $2.70 \ M_{\odot}$. Consequently, all NS solutions with central density greater than this critical value are unstable. This result is similar to GR where the change of stability also occurs at the maximum mass solution. We observe the same behavior for all of the values of the coupling constant $\alpha$ in our numerical solutions, see Fig. \ref{fig:Fundamental-Mode-SLy}. 

\begin{figure}
\centering
\subfigure[]{\includegraphics[width=0.45\linewidth]{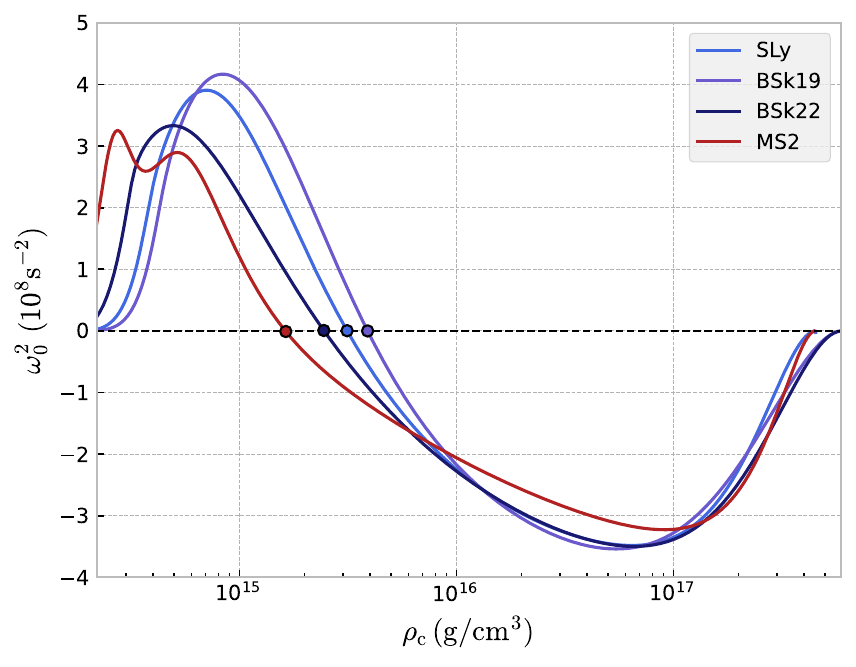}
\label{fig:freq-rho-SLy-BSk-MS2}}
\subfigure[]{\includegraphics[width=0.45\linewidth]{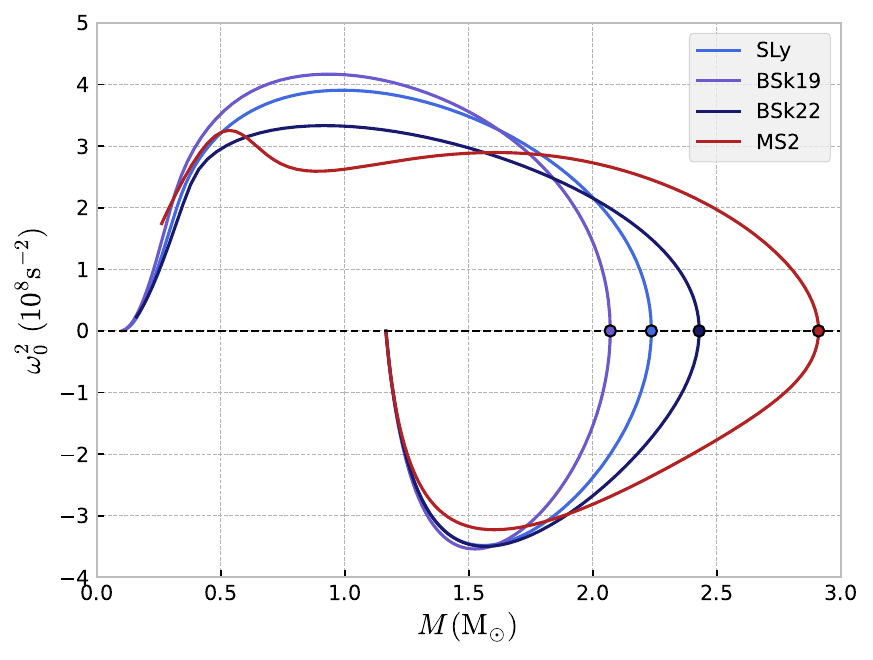}
\label{fig:freq-M-SLy-BSk-MS2}}

\caption{Fundamental eigenfrecuency versus central density (a) and mass (b) curves for neutron stars using the SLy, BSk19-22, and MS2 EOSs in 4DEGB gravity for $\alpha = 3 \ {\rm km^2}$. In the plots (a) and (b) the colored circles mark the spot where the fundamental eigenfrequency is zero, which correspond to the maximum mass solution.}
\label{fig:SLy-BSk-MS2-Freq}
\end{figure}

In Figs. \ref{fig:osc-modes} and \ref{fig:osc-modes-v2}, the results of the fundamental eigenfrequency are shown for the BSk19, BSk22, and MS2 EOSs for different values of $\alpha$. In Fig. \ref{fig:SLy-BSk-MS2-Freq}, we compare the fundamental frequencies of the SLy, BSk, and MS2 EOSs. Based on these results we can say that the change of stability for NS in 4DEGB gravity also occurs at the maximum of the mass solution (independent of the EOS or whether it is relativistic). Interestingly, in the fundamental eigenfrequency versus central density plots we notice the curves start to approach a positive value again near the BH horizon of the theory. We can interpret this as if the NS solutions were trying to return to stability near maximum central density, unlike in Einstein's theory, where they remain wholly unstable. 
We find that higher values of the 4DEGB coupling $\alpha$ tend to increase the mass of neutron stars of the same radius
 until the maximum mass point is at the end of the mass-radius curve. These interesting parts of solution space show possible ECOs, objects too dense to exist in GR. For instance, if $\alpha = 100 \ {\rm km^2}$ we obtain NS solutions that are stable under radial perturbations and  that satisfy the causality condition with masses greater than a black hole in GR for the same radius (see Figs. \ref{fig:MvsR-SLy-BSk} and \ref{fig:MvsR-MS2}).

\section{Discussion and Conclusion}\label{sec:conc}

We have investigated the stability of neutron stars in 4D Einstein-Gauss-Bonnet gravity with the SLy and BSk family as well as with the MS2 equations of state. We found that the neutron star solutions in 4DEGB gravity are qualitatively similar for each of these EOSs. We have explored the effects of different possible values of the 4DEGB coupling constant $\alpha$ and found that larger values of $\alpha$ tend to inflate the mass-radius profiles.  In Table \ref{tab:m_max}, we show the maximum masses and the corresponding radii for $\alpha = 0$ (GR),  $\alpha = 10 \ {\rm km^2}$ and  $\alpha = 100 \ {\rm km^2}$ and for the different EOSs. We see that  the maximum masses increase  with increasing $\alpha$, whereas the radii stay essentially unaltered. In addition,   differences amongst the maximum masses for different EOSs are suppressed for higher values of $\alpha$. For $\alpha = 300 \ {\rm km^2}$, $M_{\rm max} = 11.66 \ { M_{\odot}}$ and $R = 17.33 \ {\rm km}$  for all realistic EOSs considered in this work, these values approach the mass and radius of the lightest associated 4DEGB black hole.

\begin{figure}

\centering
\subfigure[MS2]{\includegraphics[width=0.45\linewidth]{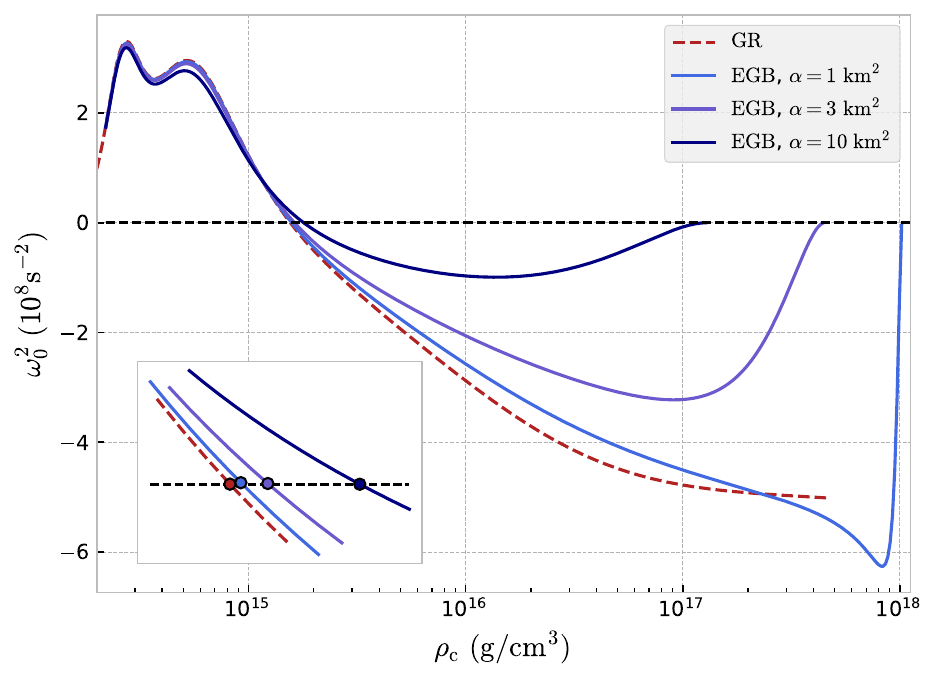}
\label{fig:Fundamental-Mode-MS2}}
\subfigure[MS2]{\includegraphics[width=0.45\linewidth]{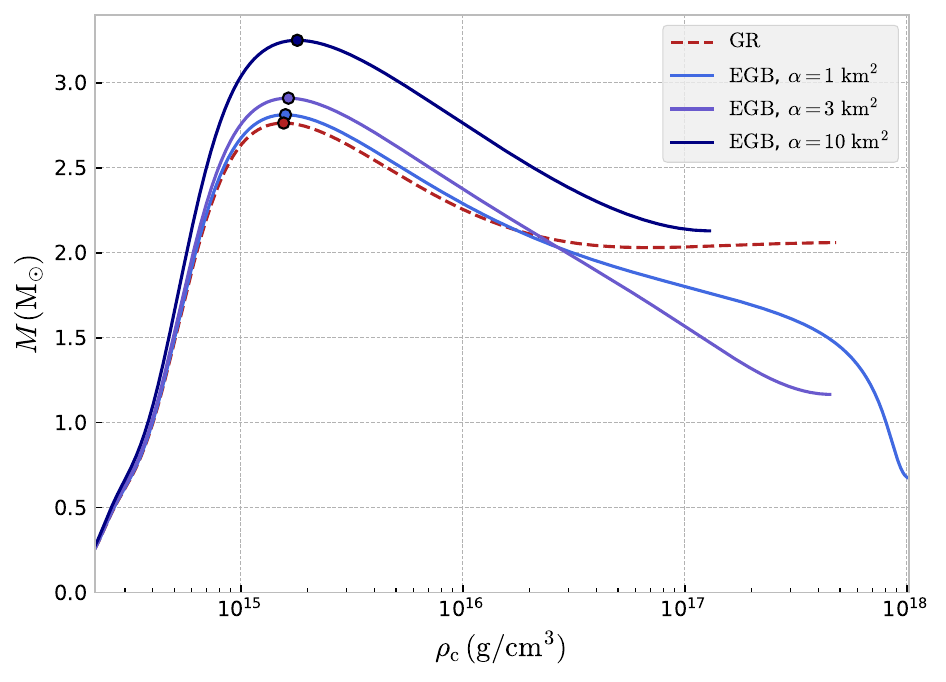}
\label{fig:M_rho-MS2}}

\caption{Fundamental eigenfrequency and mass versus central density curves for neutron stars using the MS2 EOS in GR (red dashed line) and in 4DEGB for different values of $\alpha$ (blue lines). In plot (a), the green circles mark the spot where the fundamental eigenfrequency is zero, while in plot (b) they mark the maximum mass. We note the lack of vertical lines marking the transition from a subluminal to superluminal sound speed in these plots, as the MS2 EOS   respects causality.}
\label{fig:osc-modes-v2}
\end{figure}

Our numerical results indicate that the coincidence of the maximum mass points with the transition to instability still holds in this modified theory of gravity. Rather surprisingly, we have found solutions for which the fundamental eigenfrequency $\omega_0^2$ returns to zero for large values of central density. This happens, for instance, when $\rho_{\rm c} \sim 10^{17} \ {\rm g/cm^3}$ for the MS2 EOS and $\alpha = 10 \ {\rm km^2}$, exhibited in Fig. \ref{fig:Fundamental-Mode-MS2}.  Such results hint at possible strange black hole-sized stable objects that are not present in GR - this leaves an interesting avenue for future research.  

For some EOSs and some values of $\alpha$ we found that the maximum mass points are not reached before causality-violating pressures are required; however, maximal solutions do exist that respect causality. For example, for the BSk19 EOS the maximum mass solution is not physical for $\alpha \lesssim 100 \ {\rm km^2}$, however the maximal solution for $\alpha = 300 \ {\rm km^2}$ does respect causality [see Fig. \ref{fig:Mrho-BSk19}]. For large enough coupling to the higher curvature gravity terms a maximum mass is not attained until the solution curve merges with the black hole horizon. This  means that there is a whole range of stable 4DEGB objects that are disallowed in GR, many of which are smaller than the GR Schwarzschild radius. Observation of an ECO with these characteristics could be interpreted as evidence for these higher curvature contributions to gravity having an important role to play in real gravitational dynamics.

\begin{table}[h]
    \centering

    \begin{tabular}{c|cc|cc|cc}
        \hline\hline
    &  \multicolumn{2}{c|}{$\alpha = 0$}    &     \multicolumn{2}{c|}{$\alpha = 10 \ {\rm km^2}$} & \multicolumn{2}{c}{$\alpha = 100 \ {\rm km^2}$}    \\
    EOS & $M_{\rm max} \ {(M_{\odot})}$ & $R \ {\rm (km)}$  & $M_{\rm max} \ { (M_{\odot})}$ & $R \ {\rm (km)}$ & $M_{\rm max} \ { (M_{\odot})}$ & $R \ {\rm (km)}$  \\

        \hline 
         BSk19 &    1.86         & 9.10 & 2.58        & 9.04 & 6.73 & 10.61  \\
        SLy &       2.05        &  9.98 & 2.70        & 9.97 & 6.74 & 11.34 \\
        BSk22 &     2.26        & 11.20 &  2.84        & 11.11 & 6.75 & 12.17 \\
        MS2   &     2.78        & 13.24 &    3.25        & 13.42 & 6.80 & 14.37\\
        \hline\hline
    \end{tabular}
\caption{Maximum masses with their respective radii for SLy, BSk19, BSk22, MS2 EOSs in GR and 4DEGB for $\alpha=0$ (GR) and $\alpha = 10 \ {\rm km^2}$.}
    \label{tab:m_max}
\end{table}

\section*{Acknowledgments}
This work was supported in part  by FONDECYT Grant No. 11191175 and by the Natural Sciences and Engineering Research Council of Canada.

\appendix

\section{FIELD EQUATIONS FOR THE STATIC CASE}\label{appFE}

Replacing the metric \eqref{eq:static-metric} and the components of the energy-momentum tensor \eqref{eq:Tmunu-static} in the field equations \eqref{eq:metric-FEq}, the nonredundant equations are the components $(tt)$ and $(rr)$:
\begin{gather}
    \begin{aligned} \label{eq:tt-static}
       \frac{8\pi G}{c^4} r^{2} f \epsilon &= r^{2} f^{3} \phi'^{4} \alpha +2 r^{2} f^{2} \phi'^{3} f' \alpha +4 r^{2} f^{3} \phi'^{2} \phi'' \alpha -6 r f^{2} \phi'^{2} f' \alpha -8 r f^{3} \phi' \phi'' \alpha -2 f^{3} \phi'^{2} \alpha  \\ 
        & \quad -2 f^{2} \phi'^{2} \alpha  + 6 f^{2} \phi'f' \alpha +4 f^{3} \phi'' \alpha -2 f \phi' f' \alpha -4 f^{2} \phi'' \alpha -r f f'-f^{2}+f,
    \end{aligned}
    \\
    \begin{aligned} \label{eq:rr-static}
        \frac{8\pi G}{c^4} r^{2} P &= 1 - 6 f \phi'f' \alpha +6 f^{2} \phi'^{2} \alpha +2 \phi'f' \alpha -6 f^{2} \phi'\chi' \alpha +2 f \phi' \chi' \alpha -2 f^{2} r^{2} \phi'^{3} \chi' \alpha +6 f^{2} r \phi'^{2} \chi' \alpha  \\
        &\quad -2 f r^{2} \phi'^{3} f' \alpha +6 f r \phi'^{2} f' \alpha -8 f^{2} r \phi'^{3} \alpha +3 f^{2} r^{2} \phi'^{4} \alpha -2 f \phi'^{2} \alpha +f \chi' r+f+f' r.
    \end{aligned}
\end{gather}

\section{PARAMETERS OF THE EOS} \label{app:EoS}

The parameters of the fits \eqref{eq:SLy-EoS}-\eqref{eq:MS2-EoS} are given in Table \ref{tab:SLy-parameter}-\ref{tab:MS2-parameter}, respectively.

\begin{table}[h]
    \centering
    \begin{tabular}{c|c||c|c}
    \hline \hline
      $n$ & $a_n$  & $n$ & $a_n$\\
      \hline
      1   & 6.22  & 10   & 11.4950 \\
      2   & 6.121 &   11   & $-22.775$      \\
      3   & 0.005925 &   12   & 1.5707       \\
      4   & 0.16326 &   13   & 4.3        \\
      5   & 6.48    &  14   & 14.08    \\
      6   & 11.4971 &  15   & 27.80    \\
      7   & 19.105  &  16   & $-1.653$    \\
      8   & 0.8938  &    17   & 1.50     \\
      9   & 6.54  & 18   & 14.67       \\
      \hline \hline
    \end{tabular}
    \caption{Parameters of the fit \eqref{eq:SLy-EoS}, according to Ref. \cite{Haensel2004}.}
    \label{tab:SLy-parameter}
\end{table}

\begin{table}[h]
    \centering
    \begin{tabular}{c|c c}
    \hline \hline
      $n$ & $a_n$ (BSk19) &$a_n$ (BSk22)  \\
      \hline
      1   & 3.916 & 6.682       \\
      2   & 7.701 & 5.651        \\
      3   & 0.00858 & 0.00459       \\
      4   & 0.22114 & 0.14359       \\
      5   & 3.269 & 2.681       \\
      6   & 11.964 & 11.972       \\
      7   & 13.349 & 13.993        \\
      8   & 1.3683 & 1.2904        \\
      9   & 3.254 & 2.665        \\
      10   & $-12.953$ & $-27.787$        \\
      11   & 0.9237 & 2.0140        \\
      12   & 6.20 & 4.09        \\
      13   & 14.383 & 14.135        \\
      14   & 16.693 & 28.03        \\
      15   & $-1.0514$ & $-1.921$        \\
      16   & 2.486 & 1.08        \\
      17   & 15.362 & 14.89       \\
      18   & 0.085 & 0.098       \\
      19 & 6.23 &  4.75\\
      20 & 11.68 &  11.67 \\
      21 & $-0.029$ & $-0.037$\\
      22 & 20.1 & 11.9 \\
      23 & 14.19 & 14.10 \\
      \hline \hline
    \end{tabular}
    \caption{Parameters of the fit \eqref{eq:BSk-EoS}, according to Refs. \cite{Potekhin2013,Pearson2019}.}
    \label{tab:BSk-parameter}
\end{table}

\begin{table}[h]
    \centering
    \begin{tabular}{c|c c}
    \hline \hline
      $n$ & $c_n$ & $a_n$\\
      \hline
      1   & 10.6557 & 14.0084\\
      2   & 3.7863  & 13.8422      \\
      3   & 0.8124  & 16.5970       \\
      4   & 0.6823  & $-1.0943$        \\
      5   & 3.5279  & 5.6701    \\
      6   & 11.8100 & 14.8169    \\
      7  & 12.0584  & $-56.3794$\\
      8   & 1.4663  & 9.6159\\
      9 & 3.4952    & $-0.2332$\\
      10 & 11.8007  & $-3.8369$ \\
      11 & 14.4114  & 23.1860\\
      12 & 14.4081 & ---\\
      \hline \hline
    \end{tabular}
    \caption{Parameters of the fit \eqref{eq:MS2-EoS}, according to Ref. \cite{Gungor2011}.}
    \label{tab:MS2-parameter}
\end{table}

\section{RADIAL OSCILLATIONS} \label{app:RO}

The coefficients for the dynamic equations \eqref{eq:EGB-dphi} and \eqref{eq:EGB-BC-zeta} for the radial displacement and the perturbation of the scalar field are given, respectively, by
\begin{gather}
    A_1 = \frac{(\epsilon_0 + P_0)e^{\chi_0/2}}{c^2 r^2 f_0}, \\
    A_2 =- \Gamma P_0 r^{-2} f_0 e^{3\chi_0/2}, \\
    A_3 = -\frac{d}{dr} \left[\Gamma P_0 r^{-2} f_0 e^{3\chi_0/2} \right], \\
    \begin{aligned}
    A_4 &= \frac{(\epsilon_0 + P_0)e^{3\chi_0/2}}{4r^2} \left(2f_0 \chi_0'' + 2 f_0'' - \frac{f_0'^2}{f_0}\right) \\
    & \quad + \frac{(\epsilon_0 + P_0)e^{3\chi_0/2}}{4 f_0 r^3  (r^2 - 2\alpha f_0 + 2\alpha)} \left[ - 4 \alpha  \chi_0' (r \chi_0' - 2) f_0^{3/2} - 4 \alpha r  \chi_0' f_0' \sqrt{f_0} \right. \\
    &\quad  + r  f_0^2 (r^2 + 2\alpha f_0 + 2\alpha) \chi_0'^2 - f_0 \chi_0' \left( - 2 r  (r^2 + 2\alpha) f_0' - 6 f_0^2 \alpha  \right. \\
    &\quad \left. + 4  (r^2 + 2\alpha) f_0 - \frac{8\pi G}{c^4} \epsilon_0 r^4 + \frac{8\pi G}{c^4} P_0 r^4 + 2  (r^2 + 3\alpha)\right) \\
    &\quad \left. + f_0' \left( 8f_0^2 \alpha  - 4  (r^2 + 2\alpha) f_0 + \frac{8\pi G}{c^4} r^4 (\epsilon_0 + P_0)\right)\right], 
    \end{aligned}
    \\
    A_5 =  \frac{2\alpha (\epsilon_0 + P_0) \left(r \sqrt{f_0} \chi_0' + \frac{r f_0'}{\sqrt{f_0}} - 2 \sqrt{f_0} + 2\right) e^{\chi_0}}{r^2 - 2\alpha f_0 + 2\alpha}, \\
    B_1 = 8r^4(r f_0^{2}\chi_0' + 2 f_0^{3/2} +r f_0 f_0' - 2f_0^{2}), \\
    B_2 = 4r^4(r\chi_0'^2f_0^2 + 2f_0^{3/2}\chi_0' + 4rf_0f_0'\chi_0' + 2f_0^2r\chi_0'' + 2rf_0f_0'' + f_0'^2r + 4f_0'\sqrt{f_0} - 4f_0f_0'), \\
    B_3 = -4\chi_0'r(\chi_0'f_0^2r + 2f_0^{3/2} + rf_0f_0' - 2f_0^2)e^{\chi_0/2}, \\
    \begin{aligned}
       B_4 &= -2\left(2r^2\chi_0'^3f_0^2 + 4\chi_0'^2f_0^{3/2} r + 5\chi_0'^2r^2f_0f_0' + 4r^2f_0^2\chi_0''\chi_0' - 8r\chi_0'^2f_0^2 \right. \\
       &\quad + 2\chi_0'r^2f_0''f_0 + \chi_0'r^2f_0'^2 + 4f_0^{3/2}\chi_0''r + 2r^2f_0\chi_0''f_0' - 12f_0^{3/2}\chi_0' \\
       &\quad \left.+ 4\chi_0'\sqrt{f_0}f_0'r - 10rf_0f_0'\chi_0' - 4f_0^2r\chi_0'' + 12f_0^2\chi_0'\right)e^{\chi_0/2}.  
    \end{aligned}  
\end{gather}

\bibliography{References}

\end{document}